# PrivLava: **Synthesizing Relational Data with Foreign Keys under Differential Privacy**

Kuntai Cai
caikt@comp.nus.edu.sg
National University of Singapore
Singapore, Singapore

Xiaokui Xiao
xkxiao@nus.edu.sg
National University of Singapore
Singapore, Singapore

Graham Cormode
g.cormode@warwick.ac.uk
University of Warwick
Coventry, UK

## ABSTRACT

Answering database queries while preserving privacy is an important problem that has attracted considerable research attention in recent years. A canonical approach to this problem is to use *synthetic data*. That is, we replace the input database $\mathcal{R}$ with a synthetic database $\mathcal{R}^*$ that preserves the characteristics of $\mathcal{R}$, and use $\mathcal{R}^*$ to answer queries. Existing solutions for relational data synthesis, however, either fail to provide strong privacy protection, or assume that $\mathcal{R}$ contains a single relation. In addition, it is challenging to extend the existing single-relation solutions to the case of multiple relations, because they are unable to model the complex correlations induced by the foreign keys. Therefore, multi-relational data synthesis with strong privacy guarantees is an open problem.

In this paper, we address the above open problem by proposing PrivLava, the first solution for synthesizing relational data with foreign keys under *differential privacy*, a rigorous privacy framework widely adopted in both academia and industry. The key idea of PrivLava is to model the data distribution in $\mathcal{R}$ using *graphical models*, with *latent variables* included to capture the inter-relational correlations caused by foreign keys. We show that PrivLava supports arbitrary foreign key references that form a directed acyclic graph, and is able to tackle the common case when $\mathcal{R}$ contains a mixture of public and private relations. Extensive experiments on census data sets and the TPC-H benchmark demonstrate that PrivLava significantly outperforms its competitors in terms of the accuracy of aggregate queries processed on the synthetic data.

## 1 INTRODUCTION

Answering database queries while preserving individual privacy is an important problem that has attracted considerable research effort [2, 4–6, 8, 12, 15, 22, 25, 27, 28, 33, 35, 37, 38, 40, 41, 43, 45, 46] in recent years. The bulk of existing solutions have adopted an *output perturbation* approach [35], where the database engine is modified to perturb each query answer with noise for privacy protection. The amount of noise needed is calibrated according to the amount of sensitive information that the query answer may reveal, typically based on a formal privacy framework such as *differential privacy (DP)* [16] or its generalizations.

Despite the popularity of output perturbation, it suffers one major deficiency that restricts its application in practice: the database can only answer a limited number of queries with good accuracy. This observation (formalized by Dinur and Nissim [14]) follows since each noisy answer returned by the database should inevitably reveal some information about the underlying data. As more queries are processed, the total amount of information revealed monotonically increases, and will eventually reach a threshold where answering any new queries leads to an excessive overall privacy risk, at which point the database can no longer answer new queries accurately. For this reason, existing work may focus primarily on providing a per-query privacy guarantee, and not address providing an overarching bound for all queries combined [2, 12, 35, 40, 45, 46]. Nevertheless, for a meaningful guarantee we should seek a strong privacy bound that holds for the lifetime of the database.

A natural mitigation is to build a privacy-preserving database using *synthetic data* that can be queried indefinitely without degrading the privacy with regard to the original ground truth data. Specifically, given a database $\mathcal{R}$ that contains sensitive information, we generate a synthetic version $\mathcal{R}^*$ of $\mathcal{R}$ that mimics the latter but conceals the private data therein; after that, we can use $\mathcal{R}^*$ (instead of $\mathcal{R}$) to answer queries, without imposing any constraints on the amount of queries that the users may issue. The rationale is that because all queries are processed on the synthetic database $\mathcal{R}^*$, the query answers only reveal the information in $\mathcal{R}^*$ instead of $\mathcal{R}$; therefore, as long as $\mathcal{R}^*$ itself preserves privacy, we do not need to restrict the number of queries answered using $\mathcal{R}^*$. Furthermore, in contrast to output perturbation, the usage of synthetic data $\mathcal{R}^*$ does not require any change to the database engine, nor does it incur any additional query processing overheads.

Nevertheless, to ensure strong privacy protection in $\mathcal{R}^*$, it is important that $\mathcal{R}^*$ should be generated using an algorithm that provides a rigorous privacy assurance such as DP. In relation to this, there exist a number of DP methods [7, 9, 13, 18, 21, 23, 29, 34, 36, 47, 49, 50] for synthesizing relational data. Unfortunately, all of them are designed for the restricted case when the input database $\mathcal{R}$ contains *a single relation*. In particular, they assume that the database tuples are independent samples from an unknown multi-dimensional distribution; they infer this distribution based on the observed tuples in the input table, and then sample from the inferred distribution to generate synthetic data. It is difficult to extend these methods to the general case when $\mathcal{R}$ consists of multiple relations with foreign key constraints, because the foreign keys lead to complex correlations among the tuples in different parts of the database, whereas such correlations are ignored in single-relation synthesis methods. As a consequence, synthetic data generation with foreign keys under DP is an open problem.

**Our contributions.** This paper presents PrivLava[1], the first DP solution for synthesizing relational data with foreign key constraints. The key idea of PrivLava is to model the data distribution in the input database using *graphical models* [44], with *latent variables* included to capture the inter-relational correlations caused by foreign keys. For example, consider a census database containing two relations $R_I$ and $R_H$, such that each tuple in $R_I$ (resp. $R_H$) represents an individual (resp. household), and that $R_I$ has a foreign key referencing $R_H$ to connect each individual to the household that she belongs to. To model $R_I$ and $R_H$, we assume that (i) each household is associated with a latent variable $Z$ that indicates its *type* (*e.g.*, whether it is a nuclear family), and (ii) $Z$ also decides how the household size, attributes, and members are distributed. Based on this assumption, we construct a graphical model with latent variables to infer the type $Z$ of each household based on its member composition, which then enables us to estimate the overall distribution $p(Z)$ of household types, as well as the conditional distributions of household sizes, attributes, and members, respectively, given the household type $Z$. Once we have these estimated distributions, we can use them to generate synthetic households as follows: we first sample a household type $Z$ from $p(Z)$, and then generate a synthetic household tuple by sampling its attributes based on $Z$; after that, we sample a household size $s$ given $Z$, and proceed to generate $s$ individual tuples based on $Z$. In other words, we use $Z$ to bridge the generation of the synthetic households and individuals, to preserve the distribution of household compositions induced by the foreign key between $R_I$ and $R_H$.

In general, PrivLava models each foreign key in the input database $\mathcal{R}$ separately using a graphical model with latent variables, and injects carefully calibrated noise in the model training algorithm to ensure DP. Its modeling approach is sufficiently general that it supports arbitrary foreign key references that form a directed acyclic graph. In addition, it is able to tackle the common case when $\mathcal{R}$ contains a mixture of public and private relations.

We experimentally evaluate PrivLava against the state-of-the-art DP data synthesis methods [9, 34, 49], using two census data sets from the *Integrated Public Use Microdata Series* [10, 42] as well as the TPC-H benchmark [1]. Our experimental results show that PrivLava significantly outperforms its competitors in terms of the accuracy of aggregate queries processed on the synthetic data that they generate. In addition, PrivLava is able to process a large benchmark data set in a matter of hours on a standard machine, which is reasonable for a process that would be performed once in order to release a data set.

## 2 PROBLEM DEFINITION

Let $\mathcal{R}$ be a database containing relations $R_0, R_1, R_2, \ldots$. For any $R_i, R_j \in \mathcal{R}$, we say that $R_i$ refers to $R_j$, if $R_i$ has a foreign key referencing $R_j$'s primary key. Accordingly, we say that a tuple $t_i \in R_i$ refers to a tuple $t_j \in R_j$ (denoted as $t_i \rightarrow t_j$), if the foreign key of $t_i$ equals the primary key of $t_j$. In addition, a tuple $t$ *depends on* $t_j$ (denoted as $t \rightsquigarrow t_j$), if either $t \rightarrow t_j$ or $t$ refers to another tuple that depends on $t_j$. We assume that the foreign key references in $\mathcal{R}$ form a directed acyclic graph, *i.e.*, $t \rightsquigarrow t$ never occurs.

Without loss of generality, assume that $R_0$ is a private relation that contains sensitive information. Let $t_0$ be a tuple in $R_0$, and $\mathcal{R}'$ be a database obtained by removing, from $\mathcal{R}$, $t_0$ and all other tuples that depend on $t_0$. We refer to $\mathcal{R}$ and $\mathcal{R}'$ as *neighboring databases*. Based on this notion of neighboring databases, we have the following formalization of *differential privacy*:

*Definition 2.1 (Differential Privacy (DP) [16]).* Let $F$ be an algorithm that takes as input a database. $F$ satisfies $(\epsilon, \delta)$-differential privacy (DP), if and only if for any two neighboring databases $\mathcal{R}$ and $\mathcal{R}'$ and any possible set $O$ of outputs from $F$,

$$\Pr[F(\mathcal{R}) \in O] \leq e^{\epsilon} \cdot \Pr[F(\mathcal{R}') \in O] + \delta.$$

Intuitively, the above notion of differential privacy aims to ensure that when an adversary observes the output of $F$, he is unable to infer the existence or absence of any single tuple $t$ in the relation $R_0$, even if he takes into account the dependencies among the tuples induced by foreign keys. For convenience, we refer to $R_0$ as the *primary* private relation. For any other relation $R_i$, we refer to $R_i$ as a *secondary* private relation if the tuples in $R_i$ depend on the tuples in $R_0$; otherwise, $R_i$ is a *public* relation, *i.e.*, it can be directly published without incurring any privacy cost.

Following previous work [15, 25, 43], we assume that each foreign key in the database has a bounded *multiplicity*. That is, whenever $R_i$ has a foreign key referencing $R_j$, each tuple in $R_j$ is referred to by at most $\tau_{ij}$ tuples in $R_i$, where $\tau_{ij}$ is a predefined constant. This assumption is to ensure that each tuple in the primary private relation $R_0$ is depended on by a bounded number of other tuples; otherwise, inserting or removing one tuple $t$ in $R_0$ may incur unbounded changes in the database, which makes it infeasible to enforce differential privacy since an adversary can easily infer whether $t$ exists in $R_0$. For this reason, all previous work assumes that each foreign key has a bounded multiplicity. In practice, we can enforce this assumption by *truncating* the data, *i.e.*, from each $R_j$, we remove the tuples that are referred to by more than $\tau_{ij}$ tuples in any $R_i$ [25, 43]. When each threshold $\tau_{ij}$ is reasonably large, the truncation only removes outliers in the data set, which would not significantly affect the fidelity of the data.

Our objective is to develop a synthetic data generation algorithm that (i) takes as input a database $\mathcal{R}$, (ii) outputs a database $\mathcal{R}^*$ that has the same schema as $\mathcal{R}$, and (iii) satisfies $(\epsilon, \delta)$-DP.

## 3 SOLUTION OVERVIEW

In a nutshell, PrivLava models each foreign key in the input database $\mathcal{R}$ separately using a graphical model. It combines these models together to generate synthetic relations that preserve the correlations induced by the foreign key constraints. In what follows, we present an overview of the modeling approach adopted by PrivLava, starting from the base case when $\mathcal{R}$ contains only two private relations and one foreign key.

[1] Privacy-preserving graphical models with latent variables.

**Table 1: Individual table $R_1$.**

| ID | Age | ... | H-ID |
|---|---|---|---|
| 1 | 28 | ... | 1 |
| 2 | 25 | ... | 1 |
| 3 | 27 | ... | 2 |
| 4 | 29 | ... | 2 |
| 5 | 35 | ... | 3 |
| 6 | 36 | ... | 3 |
| 7 | 5 | ... | 3 |
| ... | ... | ... | ... |

**Table 2: Household table $R_0$.**

| H-ID | ... |
|---|---|
| 1 | ... |
| 2 | ... |
| 3 | ... |
| ... | ... |

**Table 3: $R_1$ with latent $Z$.**

| ID | Age | ... | H-ID | $Z$ |
|---|---|---|---|---|
| 1 | 28 | ... | 1 | $z_1$ |
| 2 | 25 | ... | 1 | |
| 3 | 27 | ... | 2 | $z_1$ |
| 4 | 29 | ... | 2 | |
| 5 | 41 | ... | 3 | $z_2$ |
| 6 | 39 | ... | 3 | |
| 7 | 10 | ... | 3 | |
| ... | ... | ... | ... | ... |

**Table 4: $R_0$ with latent $Z$.**

| H-ID | ... | $Z$ |
|---|---|---|
| 1 | ... | $z_1$ |
| 2 | ... | $z_1$ |
| 3 | ... | $z_2$ |
| ... | ... | ... |

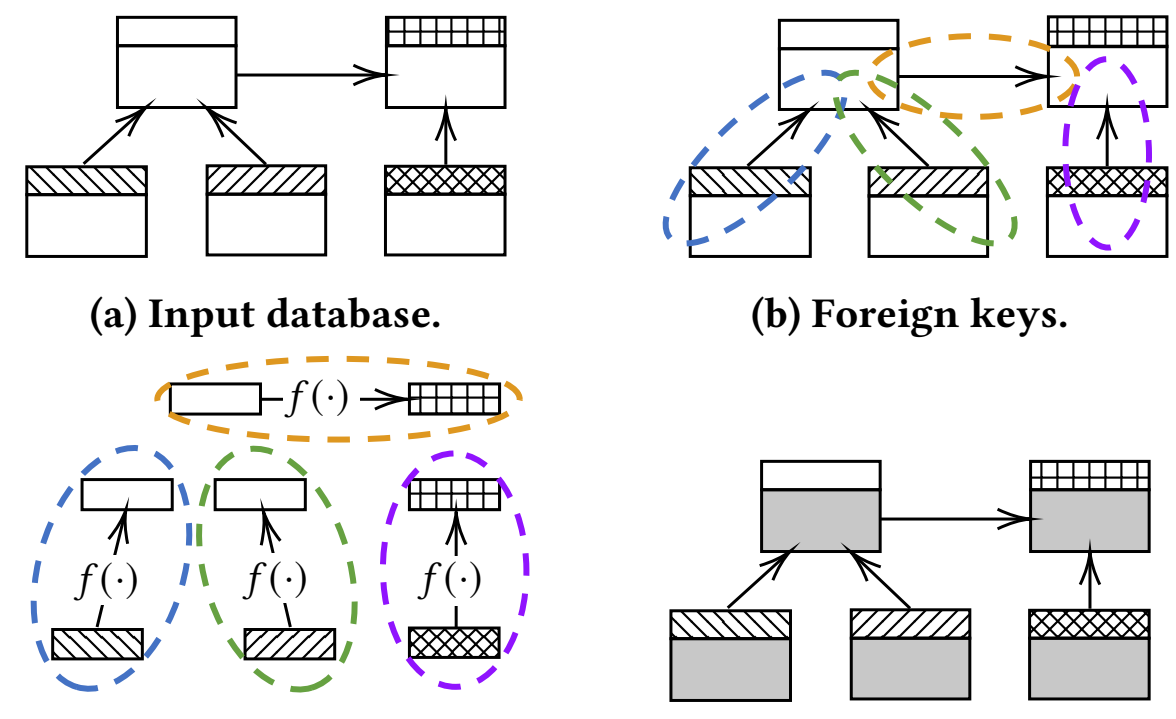

(a) Input database. (b) Foreign keys. (c) Foreign key models. (d) Synthetic database.

**Figure 1: Solution overview.**

## 3.1 Handling One Foreign Key

Assume that $\mathcal{R}$ consists of a primary private relation $R_0$ and a secondary private relation $R_1$. We say that two tuples in $R_1$ belong to the same *group*, if they refer to the same tuple in $R_0$. For instance, Tables 1 and 2 show an example when $R_0$ and $R_1$ contain information about households and individuals, respectively. Then, each group in $R_1$ is a set of individuals in the same household.

Intuitively, if we are to generate synthetic versions of $R_0$ and $R_1$, there are three types of information that we need to model:

(1) The *inter-attribute* correlations among the attributes in the same relation (*e.g.*, how individuals' ages correlate with their incomes);
(2) The *intra-group* correlations among the tuples in the same group in $R_1$ (*e.g.*, what types of individuals tend to co-exist in the same household);
(3) The *inter-relational* correlations between the tuples in $R_1$ and $R_0$ (*e.g.*, what types of individuals are likely to reside in suburban households).

Existing work on single-relation synthesis [7, 9, 29, 34, 49] has extensively studied the modeling of inter-attribute correlations. The typical approach is to (i) identify sets of attributes that are strongly correlated with each other, (ii) measure the joint-distribution of each attribute group under DP, and then (iii) use the joint-distributions obtained to generate synthetic tuples. However, there is no existing study on the preservation of intra-group and inter-relational correlations in data synthesis.

To address the above problem, our idea is to identify the *patterns* of tuple group compositions in $R_1$, and utilize them as a proxy to capture intra-group and inter-relational correlations. For example, in Table 1, the compositions of tuple groups may exhibit certain patterns, *e.g.*, young couples without kids, or middle-aged parents with children. We assume that each group's pattern is decided by a *latent* (*i.e.*, hidden) attribute $Z$ in $R_1$, as illustrated in Table 3. Observe that the first and second tuple groups in Table 3 both have $Z = z_1$, which indicates that their group compositions share the same pattern. Meanwhile, the third tuple group has $Z = z_2$, implying that its group composition differs appreciably from those of the first two groups.

Suppose that we are able to infer the value of the latent attribute $Z$ for each tuple group in $R_1$. In that case, we can use $R_1$ to estimate the joint-distribution $P_1$ of $Z$ and other attributes in $R_1$. In addition, we can associate the latent attribute of each group with the tuple in $R_0$ that the group refers to, which results in a version of $R_0$ augmented with $Z$, as illustrated in Table 4. This enables us to estimate the joint-distribution $P_0$ of $Z$ and the attributes in $R_0$. With the aforementioned joint-distributions, we can generate synthetic tuples as follows: we first sample a value $z$ from the distribution of $Z$ in the augmented $R_0$, and then generate a synthetic tuple $t^*$ for $R_0$ based on $P_0$, conditioned on $Z = z$; after that, we synthesize the tuple group in $R_1$ corresponding to $t^*$, based on $P_1$ and $Z = z$. Intuitively, if $Z$ accurately characterizes the patterns of tuple groups, then the synthetic tuples that we generate could preserve the intra-group and inter-relational correlations in the input data.

How do we infer the latent attribute $Z$ for each tuple group in $R_1$? We adopt *expectation maximization (EM)* [44], a classic approach for estimating the parameters of statistical models with latent variables. The high-level idea is that EM enables us to cluster the tuple groups in $R_1$ based on their similarities, and then we can assign one $Z$ value to each cluster to indicate that tuple groups in the same cluster follow similar patterns. In Section 4, we present the details of our EM-based algorithm for modeling one foreign key.

## 3.2 Handling Multiple Foreign Keys

For the general case when the input database $\mathcal{R}$ contains multiple foreign key references, PrivLava processes $\mathcal{R}$ following the steps illustrated in Figure 1. It first identifies the foreign keys pertinent to the private relations in $\mathcal{R}$, and then models each of them in turn using an approach similar to the algorithm outlined in Section 3.1. After that, it uses the models to generate synthetic data in a manner that preserves the correlations induced by the foreign keys.

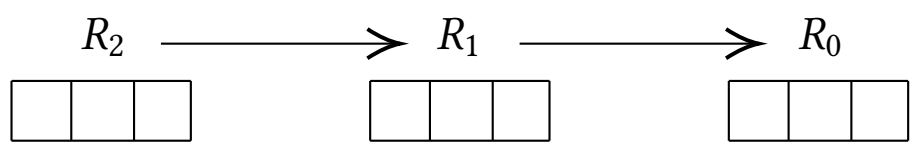

**(a) Three private relations $R_0, R_1, R_2$.**

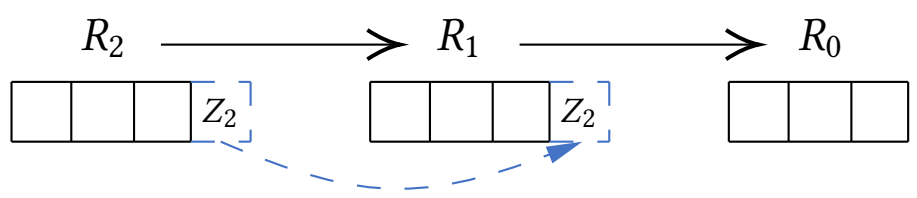

**(b) Modeling the foreign key from $R_2$ to $R_1$.**

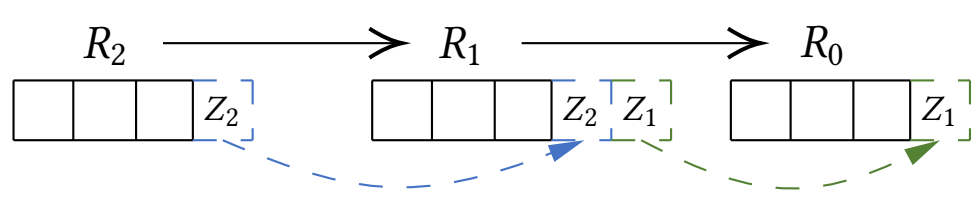

**(c) Modeling the foreign key from $R_1$ to $R_0$.**

**Figure 2: Example for the case of multiple foreign keys.**

For example, Figure 2a shows a primary private relation $R_0$ and two secondary private relations $R_1$ and $R_2$, such that $R_2$ refers to $R_1$, which in turn refers to $R_0$. Given $R_0, R_1, R_2$, PrivLava first processes the foreign key from $R_2$ to $R_1$, which leads to the augmentation of both $R_1$ and $R_2$ with a latent attribute $Z_2$. After that, PrivLava processes the foreign key from $R_1$ to $R_0$, taking $Z_2$ into account when characterizing the tuple groups in $R_1$ (*i.e.*, it treats $Z_2$ as if it is an attribute of $R_1$). This results in the augmentation of $R_1$ and $R_0$ with another latent attribute $Z_1$.

After the foreign key models are constructed, PrivLava uses them to generate synthetic data in the reverse order of foreign key modeling. In particular, given the augmented relations in Figure 2c, PrivLava first samples $Z_1$, based on which it generates a synthetic tuple in $R_0$ as well as a corresponding group $S$ of tuples in the augmented $R_1$. After that, for each synthetic tuple in $S$, PrivLava inspects its value on $Z_2$, based on which it constructs a corresponding tuple group in $R_2$. The process is repeated until we have sufficient numbers of synthetic tuples in $R_0$, $R_1$, and $R_2$.

In general, given a set of foreign key references that form a directed acyclic graph, PrivLava would model the foreign keys based on a topological order of the graph edges, and then synthesize data following the reverse order. Special consideration is given to cases when a private relation refers to a public relation or to multiple other relations, since such cases complicate the process of synthetic tuple generation. In Section 5, we elaborate our solution for handling multiple foreign keys.

## 4 SOLUTION FOR ONE FOREIGN KEY

This section presents our solution for the base case when the input database $\mathcal{R}$ contains only a primary private relation $R_0$ and a secondary private relation $R_1$, as exemplified by Tables 1 and 2. Sections 4.1-4.3 elaborate our modeling of $R_1$, and then Section 4.4 explains how we model $R_0$. After that, Section 4.5 discusses the generation of synthetic data based our modeling of $R_0$ and $R_1$.

### 4.1 Modeling the Secondary Private Relation $R_1$

We first tackle the secondary private relation $R_1$, and describe how to model the table augmented with a latent attribute via a graphical model. Without loss of generality, we assume that $R_1$ contains a primary key attribute $A_{\text{ID}}$ and a foreign key attribute $A_{\text{FK}}$ referencing $R_0$, as well as $d$ other attributes $A_1, A_2, \ldots, A_d$. As explained in Section 3.1, we assume that $R_1$ has a latent (unobservable) attribute $Z$ that characterizes the tuple groups induced by $A_{\text{FK}}$, as shown in Table 3. We model $R_1$ based on its *marginals*, which are histograms built upon subsets of the attributes in $R_1$. To explain, let $\mathcal{A}_i = \{A_{i_1}, A_{i_2}, \ldots, A_{i_k}\}$ be a $k$-size subset of $\{A_1, A_2, \ldots, A_d\}$, and $\text{span}(\mathcal{A}_i) = A_{i_1} \times A_{i_2} \times \cdots \times A_{i_k}$ be the $k$-dimensional space spanned by $A_{i_1}, A_{i_2}, \ldots A_{i_k}$. Suppose that we project the tuples in $R_1$ onto $\text{span}(\mathcal{A}_i)$. Then, for each point $(a_{i_1}, a_{i_2}, \ldots, a_{i_k}) \in \text{span}(\mathcal{A}_i)$, we have a count of the tuples $t$ in $R_1$ with $t[A_{i_j}] = a_{i_j}$ for all $j = 1, 2, \ldots, k$, where $t[A_{i_j}]$ denotes the value of $t$ on $A_{i_j}$. We refer to this set of counts as the *marginal* of $R_1$ on $\mathcal{A}_i$. For example, consider the relation $R_1$ in Table 5, where the domains of attributes $A_1$ and $A_2$ are both $\{0, 1\}$. Then, the marginal of $R_1$ on $\{A_1, A_2\}$ consists of four counts $c_{00}, c_{01}, c_{10}, c_{11}$, such that $c_{xy}$ is the number of tuples in $R_1$ with $A_1 = x$ and $A_2 = y$, as illustrated in Table 6.

**Table 5: Private relation $R_1$.**

| | $A_{\text{ID}}$ | $A_1$ | $A_2$ | $A_3$ | $A_{\text{FK}}$ |
|---|---|---|---|---|---|
| $t_1$ | 1 | 0 | 0 | 1 | 1 |
| $t_2$ | 2 | 1 | 0 | 0 | 1 |
| $t_3$ | 3 | 0 | 1 | 1 | 2 |
| $t_4$ | 4 | 1 | 0 | 1 | 2 |
| $t_5$ | 5 | 1 | 1 | 0 | 2 |

**Table 6: Marginal on $\{A_1, A_2\}$.**

| | | |
|---|---|---|
| $c_{00}$ | 1 | (for $A_1 = 0, A_2 = 0$) |
| $c_{01}$ | 1 | (for $A_1 = 0, A_2 = 1$) |
| $c_{10}$ | 2 | (for $A_1 = 1, A_2 = 0$) |
| $c_{11}$ | 1 | (for $A_1 = 1, A_2 = 1$) |

The marginal of $R_1$ on $\mathcal{A}_i$ is a representation of the joint-distribution of the attributes in $\mathcal{A}_i$. We utilize such representations in our solution, and extend them *conceptually* to take into account the latent attribute $Z$. In particular, for any tuple $t$ in $R_1$ and any element $z$ in the domain of $Z$, let $p(z \mid t)$ denote the probability that $t$'s tuple group has $Z = z$. (Note that this probability is not directly observable from $R_1$, since $Z$ is latent.) We consider the $(k+1)$-dimensional space $\text{span}(\mathcal{A}_i \cup \{Z\})$ spanned by $Z$ and the attributes in $\mathcal{A}_i$. For each point $(a_{i_1}, \ldots, a_{i_k}, z) \in \text{span}(\mathcal{A}_i \cup \{Z\})$, we take the sum of all $p(z \mid t)$ where $t \in R_1$ and $t[A_{i_j}] = a_{i_j}$ for $j = 1, 2, \ldots, k$. In other words, we count the expected number of tuples $t$ with $t[A_{i_j}] = a_{i_j}$ and belong to a tuple group with $Z = z$. This results in a set of $|\text{span}(\mathcal{A}_i \cup \{Z\})|$ expected counts, which we define as the marginal of $R_1$ on $\mathcal{A}_i \cup \{Z\}$. We refer to such a marginal as a *latent* marginal, and refer to a marginal defined on some $\mathcal{A}_i$ as an *observed* marginal.

Suppose that we are given a set $\mathcal{M}$ of marginals of $R_1$. We model $R_1$ using an *exponential family graphical model* [44]. The model is parameterized by a vector $\theta$, where each coordinate is associated with a count (or expected count) $c$ in a marginal in $\mathcal{M}$ (that is, the size of $\theta$ equals the total number of counts in the marginals in $\mathcal{M}$). For convenience, we use $\theta[c]$ to denote the coordinate in $\theta$ associated with $c$. Let $t$ be a tuple and $z$ be a value in the domain of $Z$, and $p(t, z)$ be the probability that if we sample a tuple from $R_1$, it would have the same attribute values as $t$ and belong to a tuple group with $Z = z$. Our graphical model assumes that $p(t, z)$ can be derived from $\theta$ based on the marginal counts (or expected counts) $c$ that are *linked to* $(t, z)$, denoted as $c \leftrightarrow (t, z)$:

$$p(t, z) \propto \prod_{c \in M \in \mathcal{M} \wedge c \leftrightarrow (t,z)} \exp(\theta[c]). \quad (1)$$

Specifically, in a marginal $M$ on an attribute set $\mathcal{A}_i$, a count $c$ is linked to $(t, z)$ if $c$ corresponds to a point $(a_{i_1}, \ldots, a_{i_k}) \in \text{span}(\mathcal{A}_i)$ with $a_{i_j} = t[A_{i_j}]$ for $j = 1, \ldots, k$. For example, in the marginal in Table 6, the count $c_{10}$ is linked to $t_2$ and $t_4$ in Table 5, since $c_{10}$ corresponds to the point $(A_1 = 1, A_2 = 0)$, while both $t_2$ and $t_4$ have $A_1 = 1$ and $A_2 = 0$. Similarly, in a marginal on an attribute set $\mathcal{A}_i \cup \{Z\}$, we say that an expected count is linked to $(t, z)$ if it corresponds to a point $(t[A_{i_1}], \ldots, t[A_{i_k}], z) \in \text{span}(\mathcal{A}_i \cup \{Z\})$.

In addition to Equation 1, we also assume that the latent attributes $z$ of the tuple groups in $R_1$ follow a distribution $p_Z$, and that the sizes of the tuple groups follow a distribution $p_{\text{size}}$ conditioned on $z$. We consider that each tuple group $G$ in $R_1$ has the following generative process:

(1) Sample a value $z$ of $Z$ with probability $p_Z(z)$;
(2) Sample a group size $s$ with probability $p_{\text{size}}(s \mid z)$;
(3) Sample $s$ tuples $t_1, t_2, \ldots, t_s$, each with probability $p(t_j \mid z) = \frac{p(t_j, z)}{\sum_t p(t, z)}$, $j = 1, 2, \ldots, s$.

Based on this generative process, for each tuple group $G$ in $R_1$, we formulate its likelihood as

$$p(G) = \sum_z \left( p_Z(z) \cdot p_{\text{size}}(|G| \mid z) \cdot \prod_{t \in G} p(t \mid z) \right). \tag{2}$$

Our objective is to estimate $\theta$, $p_Z$, and $p_{size}$ to maximize the overall log-likelihood of all tuple groups $G$ observed in $R_1$, *i.e.*, $\sum_G \log p(G)$ (note that this is equivalent to maximizing $\prod_G p(G)$). In Section 4.2, we introduce an algorithm for estimating $\theta$, $p_Z$, and $p_{\text{size}}$ when given a set $\mathcal{M}$ of marginals of $R_1$. After that, in Section 4.3, we explain how we determine $\mathcal{M}$ in our solution.

## 4.2 Estimating Parameters $\theta$, $p_Z$, and $p_{size}$

Next, we assume that a set $\mathcal{M}$ of marginals is given, and describe an algorithm for learning the parameters of the model. Suppose that $\mathcal{M}$ contains both observed and latent marginals. We assume that the counts in the observed marginals have been computed from $R_1$ and injected with Gaussian noise [3] to achieve DP. (We explain the details of noise injection in Section 4.3.) Meanwhile, for the latent marginals, we treat each of their counts as a variable, since the values of $Z$ in $R_1$ are unknown in advance. We use $\mathcal{M}$ to estimate $\theta$, $p_Z$, and $p_{\text{size}}$ using an algorithm based on the *expectation-maximization (EM)* method [44], as shown in Algorithm 1.

Algorithm 1 starts by taking initial model parameters $\theta, p_Z, p_{\text{size}}$ from the input. (We explain the initialization of these parameters in Section 4.3.) The next part of Algorithm 1 consists of $T$ iterations (Lines 1-23), each of which follows the typical framework of the EM method and consists of two steps as follows:

(1) *The E step* (Lines 2-7): Estimate the latent attribute of each tuple group $R_1$, based on the current model parameters;
(2) *The M step* (Lines 8-23): Update the model parameters to optimize the overall log-likelihood of all tuple groups $G$, *i.e.*, $\sum_G \log p(G)$.

In other words, the algorithm iteratively refines the estimation of each tuple group's latent attribute, and utilizes them to improve the accuracy of the model parameters.

Specifically, in each iteration, Algorithm 1 first inspects the tuple groups in $R_1$ (Lines 2-7). For each tuple group $G$, it computes the

**Algorithm 1:** Parameter estimation

**Input:** Relation $R_1$, initial model parameters $\theta$, $p_Z$, and $p_{\text{size}}$, iteration number $T$, marginal set $\mathcal{M}$, noise scales $\sigma_z$, $\sigma_{\text{size}}$, and $\sigma_\ell$
**Output:** Updated model parameters $\theta$, $p_Z$, $p_{\text{size}}$, latent attribute $z_G$ for each tuple group $G$

1 **for** $j = 1$ **to** $T$ **do**
2   **for** *each tuple group $G$ in $R_1$* **do**
3     **for** *each tuple $t$ in $G$ and each $z$* **do**
4       Compute $p(t \mid z) = \frac{p(t,z)}{\sum_{t'} p(t',z)}$ based on $\theta$;
5     **for** *each $z$* **do**
6       Compute $p(z \mid G) = \frac{p(G,z)}{\sum_{z'} p(G,z')}$, where $p(G,z) = p_Z(z) \cdot p_{\text{size}}(|G| \mid z) \cdot \prod_{t \in G} p(t \mid z)$;
7     Let $z_G$ be the $z$ that maximizes $p(z \mid G)$;
8   **for** *each $z$* **do**
9     Count the number $\text{cnt}(z)$ of tuple groups $G$ in $R_1$ with $z_G = z$;
10     Let $\text{cnt}^*(z) = \max\left\{0, \text{cnt}(z) + \mathcal{N}(0, \sigma_z^2)\right\}$;
11   **for** *each $z$* **do**
12     $p_Z(z) = \frac{\text{cnt}^*(z)}{\sum_{z'} \text{cnt}^*(z')}$;
13   **for** *each $z$* **do**
14     **for** *each group size $s$* **do**
15       Count the number $\text{cnt}(s, z)$ of tuple groups $G$ with $z_G = z$ and $|G| = s$;
16       Let $\text{cnt}^*(s, z) = \max\left\{0, \text{cnt}(s, z) + \mathcal{N}(0, \sigma_{\text{size}}^2)\right\}$;
17     **for** *each group size $s$* **do**
18       $p_{\text{size}}(s \mid z) = \frac{\text{cnt}^*(s,z)}{\sum_{s'} \text{cnt}^*(s',z)}$
19   **for** *each latent marginal $M \in \mathcal{M}$* **do**
20     Compute the expected counts in $M$ based on the $z_G$ of each group $G$;
21     **for** *each expected count $c$ in $M$* **do**
22       $c = c + \mathcal{N}(0, \sigma_\ell^2)$;
23   Update $\theta$ based on the noisy counts and expected counts currently in the marginals in $\mathcal{M}$;
24 **return** $\theta, p_{\text{size}}$, and $z_G$ for each tuple group $G$;

conditional probability $p(t \mid z) = \frac{p(t,z)}{\sum_{t'} p(t',z)}$ for each $t \in G$ and $z$ based on the current $\theta$. Then, based on $p(t \mid z)$ and the current $p_Z$ and $p_{\text{size}}$, it derive the conditional probability

$$p(z \mid G) = \frac{p(G, z)}{\sum_z' p(G, z')}, \tag{3}$$

where

$$p(G, z) = p_Z(z) \cdot p_{\text{size}}(|G| \mid z) \cdot \prod_{t \in G} p(t \mid z). \tag{4}$$

Based on $p(z \mid G)$, it identifies the latent value $z_G$ that is most likely for $G$, and takes it as the current estimate of $G$'s latent attribute.

After that, for each $z$, the algorithm counts the number $\text{cnt}(z)$ of tuple groups whose latent attributes are estimated to be $z$, and obtains a noisy version $\text{cnt}^*(z)$ of the number by injecting Gaussian noise $\mathcal{N}(0, \sigma_z^2)$ (Lines 8-10). Then, it updates the estimated distribution $p_Z$ of $Z$ based on $\text{cnt}^*(z)$ (Lines 11-12). Similarly, for each $z$ and each group size $s$, the algorithm computes a noisy version $\text{cnt}^*(s, z)$ of the number of tuple groups $G$ in $R_1$ that have $z_G = z$ and $|G| = s$ (Lines 13-16). It then uses $\text{cnt}^*(s, z)$ to update the estimated distribution $p_{\text{size}}(s \mid z)$ of the group size $s$ given $z$.

In addition, for each latent marginal $M \in \mathcal{M}$, the algorithm calculates the expected counts in $M$ based on the estimated latent attribute of each tuple group, and then injects Gaussian noise $\mathcal{N}(0, \sigma_\ell^2)$ into them. In other words, it materializes a noisy version of each latent marginal in $\mathcal{M}$. Then, it treats all latent marginals as observed marginals (since they have been materialized), and updates the graphical model parameter $\theta$ to fit the noisy counts in all marginals in $\mathcal{M}$ (Line 23). Inferring $\theta$ from observed marginals is well studied in the literature of graphical models [44]; for Line 23 of Algorithm 1, we adopt the gradient descent method in [34], as it is optimized for the case when the marginal counts are noisy. One issue here is that the method in [34] is designed for an objective function different from ours, so it is not immediately clear that the method fits our EM framework; however, we prove in Appendix A that the solution returned by the method is still correct in our context, *i.e.*, it optimizes the overall log-likelihood of all tuple groups.

## 4.3 Choosing the Marginal Set $\mathcal{M}$

Now we remove the assumption that the marginal set $\mathcal{M}$ is given, and describe a procedure to choose the observed and latent marginals in $\mathcal{M}$. Intuitively, the observed and latent marginals in $\mathcal{M}$ serve different purposes: the observed marginals help us capture the inter-attribute correlations in $R_1$, while the latent marginals enable us to model the intra-group correlations in $R_1$. In relation to this, existing work on single-table synthesis [7, 9, 49] has proposed a number of DP solutions for choosing observed marginals to model inter-attribute correlations. Nevertheless, the selection of latent marginals under DP has not been studied.

The main challenge in latent marginal selection is the following dilemma. To choose latent marginals based on their usefulness, we need to inspect the joint-distribution of the latent attribute $Z$ and the other attributes pertinent to each marginal; this requires us to estimate the latent attribute of each tuple group. However, to estimate each tuple's latent attribute using Algorithm 1, we need to provide $\mathcal{M}$ as an input to the algorithm, which requires that we have decided the latent marginals in $\mathcal{M}$.

To resolve the above dilemma, we adopt an iterative approach as follows. We first select an initial set of latent marginals, and insert them into $\mathcal{M}$ along with a number of observed marginals. After that, we invoke Algorithm 1 to estimate our model parameters based on the current $\mathcal{M}$. With the parameters obtained, we use our model to estimate the latent attribute of each tuple group. Then, we utilize the estimated latent attributes to evaluate the usefulness of a candidate set $C$ of latent marginals currently not in $\mathcal{M}$, based on which we choose the most promising latent marginals from $C$ and insert them into $\mathcal{M}$. This process is repeated until we have a sufficient set of latent marginals in $\mathcal{M}$.

Algorithm 2 shows the pseudo-code of our method for deciding $\mathcal{M}$. It first invokes the DP marginal selection algorithm in existing

**Algorithm 2:** Marginal selection and model constrution for $R_1$

**Input:** Relation $R_1$, candidate marginal number $n_C$, marginal number increment $n_{\text{inc}}$, iteration numbers $T_C$ and $T$, noise scales $\sigma_z$, $\sigma_{\text{size}}$, $\sigma_\ell$, and $\sigma_M$

**Output:** Marginal set $\mathcal{M}$, model parameters $\theta, p_{\text{size}}$, and latent attribute $z_G$ for each tuple group $G$

1 Invoke the DP marginal selection algorithm in [9] on $R_1$ to obtain a set $\mathcal{M}_{\text{obs}}$ of observed marginals with Gaussian noise injected;
2 Let $\mathcal{M} = \mathcal{M}_{\text{obs}}$;
3 **for** *each attribute* $A \in \{A_1, A_2, \ldots, A_d\}$ **do**
4     Let $M$ be the latent marginal of $R_1$ on $\{A, Z\}$;
5     Insert $M$ into $\mathcal{M}$;
6 Initialize model parameters $\theta, p_Z, p_{\text{size}}$;
7 $\theta, p_{\text{size}}, z_G \longleftarrow$ Alg. 1 $(R_1, \theta, p_Z, p_{\text{size}}, T, \mathcal{M}, \sigma_z, \sigma_{\text{size}}, \sigma_\ell)$;
8 **for** $t = 1$ **to** $T_C$ **do**
9     Sample a set $C$ of $n_C$ latent marginals that satisfy $\lambda$-usefulness and are not in $\mathcal{M}$;
10     Let $C'$ be the observed marginal set corresponding to $C$;
11     **for** *each observed marginal* $M' \in C'$ **do**
12         Generate an estimated version $\widetilde{M'}$ of $M'$ by replacing each count with an estimate derived using the current $\theta$;
13         Let $\widetilde{\text{err}}(M') = \|M' - \widetilde{M'}\|_1 + \mathcal{N}(0, \sigma_{\text{err}}^2)$;
14     Identify the $n_{\text{inc}}$ marginals $M'$ in $C'$ that maximizes $\widetilde{\text{err}}(M')$, and insert their corresponding latent marginals into $\mathcal{M}$;
15     Update $\theta$ by inserting an new coordinate $\theta[c] = 0$ for each count $c$ in each latent marginal newly inserted into $\mathcal{M}$;
16     $\theta, p_{\text{size}}, z_G \longleftarrow$ Alg. 1 $(R_1, \theta, \mathcal{M}, p_Z, p_{\text{size}}, 1, \sigma_z, \sigma_{\text{size}}, \sigma_\ell)$;
17 **return** $\mathcal{M}$, $\theta$, $p_{\text{size}}$, and $z_G$ for each tuple group $G$;

work [9] to identify a set of observed marginals that preserve inter-attribute correlations $R_1$ and has Gaussian noise injected (Line 1). It then uses this marginal set as the initial $\mathcal{M}$ (Line 2). After that, for each attribute $A \in \{A_1, \ldots, A_d\}$, the algorithm inserts a latent marginal of $R_1$ on $\{A, Z\}$ into $\mathcal{M}$ (Lines 3-4). The rationale is that these latent marginals represent the most basic type of joint-distributions between the latent attribute $Z$ and the others, and hence, they are a good set of latent marginals to start with.

Then, the algorithm proceeds to initialize our model parameters $\theta$, $p_Z$, and $p_{\text{size}}$ (Line 6), by setting $\theta$ to a zero vector, $p_Z$ to a uniform distribution over the predefined domain of $Z$, and $p_{\text{size}}$ to a uniform distribution over $\{1, 2, \ldots, \tau\}$, where $\tau$ is the maximum multiplicity of $R_1$'s foreign key to $R_0$ (see Section 2). After that, the algorithm invokes Algorithm 1 to learn $\theta$, $p_Z$, and $p_{\text{size}}$ based on the current marginal set $\mathcal{M}$ (Line 7). The subsequent part of the algorithm consists of $T_C$ iterations (Lines 8-16), each of which identifies $n_{\text{inc}}$ additional latent marginals and inserts them into $\mathcal{M}$. In particular, in each iteration, the algorithm samples a set $C$ of $n_C$ latent marginals $M$ that are not in $\mathcal{M}$ but satisfy a property of *$\lambda$-usefulness* [49]:

$$\frac{\tilde{n}}{|M|} \geq \lambda \cdot \sqrt{\frac{2}{\pi}} \cdot \sigma_\ell, \quad (5)$$

where $\tilde{n}$ is the noisy number of tuples in $R_1$ obtained from the selection algorithm in Line 1, $|M|$ is the number of counts in $M$, $\lambda > 1$ is a pre-defined constant, and $\sqrt{\frac{2}{\pi}}\sigma_\ell$ is the expected absolute value of the Gaussian noise that Algorithm 1 will insert into $M$ if it is selected into $\mathcal{M}$ (see Line 16). Intuitively, $\lambda$-usefulness requires that the average count in $M$ is larger than the expected amount of noise to be inserted into $M$, which ensures that the signal-to-noise ratio in $M$ is reasonable after noise injection. We set $\lambda = 20$.

Next, Algorithm 2 constructs a set $C'$ of observed marginals *corresponding to* the latent marginals in $C$ (Line 10). Specifically, we say that an observed marginal $M'$ corresponds to a latent marginal $M$, if $M'$ is defined on an attribute set $\mathcal{A} \subseteq \{A_1, A_2, \ldots, A_d\}$ and $M$ is defined on $\mathcal{A} \cup \{Z\}$. Subsequently, for each observed marginal $M'$ in $C'$, Algorithm 2 evaluates how accurate the current model can approximate the counts in $M'$. In particular, the algorithm first utilizes the current model parameter $\theta$ to estimate the counts in $M'$ (based on Equation 1), obtaining an estimated version $\widetilde{M'}$ of $M'$ (Line 12). After that, it computes the $L_1$ distance between $M'$ and $\widetilde{M'}$, and injects Gaussian noise $\mathcal{N}(0, \sigma^2_{\text{err}})$ into it (Line 13). Let $\widetilde{\text{err}}(M')$ denote the resulting noisy $L_1$ distance. Intuitively, if $\widetilde{\text{err}}(M')$ is large, then the information in $M'$ is not accurately captured by our model; in that case, it is beneficial to add a marginal corresponding to $M'$ into $\mathcal{M}$, so as to improve our model. Accordingly, Algorithm 2 identifies the $n_{\text{inc}}$ marginals $M'$ in $C'$ with the largest $\widetilde{\text{err}}(M')$, and inserts their corresponding latent marginals into $\mathcal{M}$ (Line 14). After that, the algorithm updates the model parameter $\theta$ by creating a new coordinate $\theta[c] = 0$ for each cell in each newly added marginal (Line 15). Then, the algorithm invokes Algorithm 1 to update $\theta$ based on the current $\mathcal{M}$ (Line 16). Note that the iteration number in Algorithm 1 is set to $T = 1$, because the learning of model parameters here is based on the previously obtained $\theta$, $p_Z$, $p_{\text{size}}$ (instead of starting from scratch), and hence, one iteration is sufficient for fine-tuning our model.

## 4.4 Modeling the Primary Private Relation $R_0$

To complete the modeling step, we model the primary private relation $R_0$, using existing techniques. Recall that, with Algorithms 1 and 2, we obtain a DP graphical model for the secondary private relation $R_1$, as well as the inferred latent attribute $z_G$ for each tuple group $G$ in $R_1$. Then, we can insert a new attribute $Z$ into the primary private relation $R_0$ as follows: if a tuple group $G$ in $R_1$ refers to a tuple $t$ in $R_0$, then we set $t[Z] = z_G$, as exemplified in Table 4. In that case, $Z$ can be regarded as an observed attribute (instead of a latent attribute), since its value for each tuple in $R_0$ is determined. Therefore, to model $R_0$, we can apply any existing DP method for single-relation synthesis. In our solution, we model $R_0$ using the state-of-the-art approach in [9].

## 4.5 Synthesizing $R_0$ and $R_1$

Now that we have modeled the foreign key between $R_0$ and $R_1$, we discuss how to generate synthetic data from the model by a random sampling procedure. This relies on standard ideas, and can be done efficiently. First, we use the model for $R_0$ to generate a synthetic tuple $r^*$, which contains all attributes originally in $R_0$ as well as the additional attribute $Z$. After that, we retrieve the model for $R_1$, which is parameterized by $\theta$, $p_Z$, and $p_{\text{size}}$. We first inspect $r^*[Z]$, and sample a tuple group size $s$ with probability $p_{\text{size}}(s \mid z = r^*[Z])$. Then, we generate a group $G$ of $s$ tuples for $R_1$. Each tuple $t^*$ in $G$ is sampled with probability $p(t^* \mid z = r^*[Z])$, where $p(t \mid z)$ is as defined in Section 4.1. This sampling of tuples can be conducted efficiently using the *junction tree representation* of our graphical model [44], which is a standard approach adopted in previous work [9]. We omit the details for brevity. Once the tuples in $G$ are constructed, we set their foreign keys to make them refer to $r^*$. In general, the above generation process can be repeated to create an arbitrary number of tuples for $R_0$ and the corresponding tuple groups in $R_1$.

## 4.6 Extension to Multiple Latent Attributes

Our discussions in Sections 4.1-4.5 assume that we use only one latent attribute $Z$ in the modeling of $R_1$ and $R_0$. In general, however, we can have multiple latent attributes $Z_1, Z_2, \ldots$, and treat them as a composite attribute to model tuple groups, in a way similar to the case of a single latent attribute $Z$. The main advantage of using multiple latent attributes is that it provides additional flexibility in the choices of marginals in our model, since each observed attribute can be combined with different latent attributes to form latent marginals. Nevertheless, having more latent attributes also increases the difficulty of modeling, as it requires us to deal with higher-dimensional data in both $R_1$ and $R_0$.

In our solution, we choose to use two latent attributes $Z_1$ and $Z_2$ for each foreign key, as it leads to better empirical results in general. Accordingly, we revise Algorithms 1 and 2 as follows:

(1) In Algorithm 1, we use the composite attribute $\{Z_1, Z_2\}$ in place of $Z$.
(2) In Lines 3-5 of Algorithm 2, we consider the latent marginals that are defined on $\{A, Z_1\}$, $\{A, Z_2\}$, or $\{Z_1, Z_2\}$, where $A \in \{A_1, \ldots, A_d\}$.
(3) In Line 9 of Algorithm 2, we allow each latent marginal to contain $Z_1$ or $Z_2$ or both.

## 4.7 Choice of Latent Attribute Domain

Our solution requires that each latent attribute $Z_i$ is a categorical attribute with a given domain. While the elements of $Z_i$'s domain can be arbitrary, the size of the domain $|Z_i|$ has a considerable effect on the performance of our solution. To explain, recall that our solution uses $Z_i$ to characterize the composition of each tuple group in $R_1$, *i.e.*, each $Z_i$ value represents a *type* of tuple groups. When each $|Z_i|$ is small, we would have an insufficient number of types to categorize the tuple groups in $R_1$, which leads to inaccurate modeling of the data. On the other hand, if $|Z_i|$ is large, then each latent marginal would contain an excessive number of counts; in that case, the "signal" in the latent marginals is too sparse to satisfy our $\lambda$-usefulness requirement (see Line 9 of Algorithm 2), namely, we are unable to utilize the latent marginals to model $R_1$.

To address the above issue, we set the domain sizes of our latent attributes $Z_1$ and $Z_2$ as follows. Let $A$ be the attribute in $R_1$ with the largest domain. We let $|Z_1| = |Z_2| = k$, where $k$ is the largest

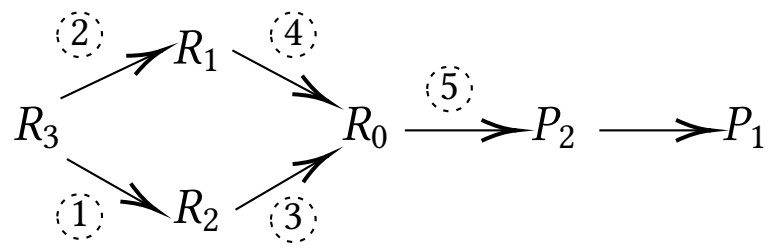

**Figure 3: An example of multiple foreign key references.**

**Algorithm 3:** Modeling multiple foreign keys

**Input:** Database $\mathcal{R}$, strict total order $\prec$

**for** *each private foreign key* $\mathrm{FK}(R, R')$ *in ascending order of* $\prec$ **do**
- Apply Algorithm 2 to model $\mathrm{FK}(R, R')$, and augment $R'$ with the estimated latent attributes;

**for** *each relation R that has no private foreign keys but is referred to by at least one private relation* **do**
- Construct a DP single-relation model for $R$ (see Section 4.4);

**return** all models constructed;

number such that the latent marginals on $\{Z_1, Z_2\}$ and $\{Z_i, A\}$ all satisfy $\lambda$-usefulness. That is, we set the domain size of $Z_i$ as large as possible, while ensuring that our solution is able to use latent marginals to model the joint-distribution between $Z_1$ and $Z_2$, as well as the correlations between each observed attribute and $Z_i$.

## 5 SOLUTION FOR MULTIPLE FOREIGN KEYS

Assume that the input database $\mathcal{R}$ contains a primary private relation $R_0$, secondary private relations $R_1, R_2, \ldots$, and public relations $P_1, P_2, \ldots$. For any two relations $R, R' \in \mathcal{R}$ such that $R$ refers to $R'$, we use $\mathrm{FK}(R, R')$ to denote the foreign key from $R$ to $R'$. By our problem definition, the foreign keys between public relations do not incur any privacy issues. In addition, there should be no foreign key from a public relation $P$ to a private relation; otherwise, $P$ should be a secondary private relation instead. Therefore, there are only two types of foreign keys in $\mathcal{R}$ that we need to tackle: (i) $\mathrm{FK}(R_i, R_j)$, where $R_i$ and $R_j$ are both private relations, and (ii) $\mathrm{FK}(R_i, P)$, where $P$ is a public relation. We refer to these two types of foreign keys as *private* foreign keys.

Observe that there exists a strict total order $\prec$ on the set of private foreign keys in $\mathcal{R}$, such that $\mathrm{FK}(R, R') \prec \mathrm{FK}(R', R'')$ always holds. For example, Figure 3 shows the foreign key references in a database with four private relations $R_0, \ldots, R_3$ and two public relations $P_1, P_2$. The circled number associated with each private foreign key indicates the index of the foreign key in a total order $\prec$. Observe that both $\mathrm{FK}(R_1, R_0) \prec \mathrm{FK}(R_0, P_2)$ and $\mathrm{FK}(R_2, R_0) \prec \mathrm{FK}(R_0, P_2)$. In general, such a strict total order can be constructed by applying a topological sort on the directed acyclic graph that represents the foreign key references.

In the following, we explain how we can extend our solution in Section 4 to address the case of multiple foreign keys, based on the aforementioned total order $\prec$.

### 5.1 Modeling Multiple Foreign Keys

Algorithm 3 shows the pseudo-code of our method for modeling $\mathcal{R}$. The algorithm performs a linear scan of the private foreign keys in $\mathcal{R}$ in ascending order of $\prec$. For each private foreign key $\mathrm{FK}(R, R')$, the algorithm applies Algorithm 2 to model $\mathrm{FK}(R, R')$, which also augments $R'$ with the estimated latent attributes associated with each tuple group in $R$. This augmented $R'$ is then used in place of the original $R'$ in the subsequent processing of any foreign key $\mathrm{FK}(R', R'')$ encountered. When the linear scan terminates, the algorithm proceeds to examine the relations in $\mathcal{R}$ that (i) have no private foreign keys but (ii) are referred to by at least one private relation. Such relations are not modeled by Algorithm 2, but they are augmented with latent attributes, which are needed in data synthesis. We model such relations using the DP single-relation modeling approach in [9], following the procedure in Section 4.4.

For example, given the database $\mathcal{R}$ in Figure 3, Algorithm 3 first processes $\mathrm{FK}(R_3, R_2)$ and augments $R_2$ with estimated latent attributes. After that, it proceeds to model $\mathrm{FK}(R_3, R_1)$, $\mathrm{FK}(R_2, R_0)$, and $\mathrm{FK}(R_1, R_0)$, respectively, and inserts estimated latent attributes into $R_1$ and $R_0$. Note that at this stage, $R_0$ is augmented with two sets of latent attributes, induced by $\mathrm{FK}(R_2, R_0)$ and $\mathrm{FK}(R_1, R_0)$, respectively. Both sets of latent attributes are taken into account when the algorithm subsequently processes $\mathrm{FK}(R_0, P_2)$.

**Algorithm 4:** Synthesis without public relations

**Input:** Database $\mathcal{R}$ without public relations, strict total order $\prec$

**Output:** A synthetic version $\mathcal{R}^*$ of $\mathcal{R}$

Synthesize $R_0^*$ from the single-relation model for $R_0$;

**for** *each private foreign key* $\mathrm{FK}(R_i, R_j)$ *in descending order of* $\prec$ **do**
- **if** $R_i^*$ *has not been generated* **then**
  - $R_i^* = \emptyset$;
  - **for** *each synthetic tuple* $t^*$ *in* $R_j^*$ **do**
    - Sample a tuple group $G^*$ for $t^*$ using the graphical model for $\mathrm{FK}(R_i, R_j)$;
    - Insert the tuples in $G^*$ into $R_i^*$;
- **else**
  - Let $S = R_i^*$;
  - **for** *each synthetic tuple* $t^*$ *in* $R_j^*$ **do**
    - Sample a tuple group $G^*$ for $t^*$ using the graphical model for $\mathrm{FK}(R_i, R_j)$, but restrict the sample space to $S$;
    - **for** *each tuple* $t_i^*$ *in* $R_i^*$ *that is also in* $G^*$ **do**
      - Let $t_i^*$ refer to $t^*$;
      - Remove $t_i^*$ from $S$;

**return** all synthetic relations generated;

### 5.2 Synthesis without Public Relations

In what follows, we explain how we generate a synthetic version of $\mathcal{R}$ based on the models constructed by Algorithm 3. For ease of exposition, we first consider the case when $\mathcal{R}$ does not contain any public relations. In that case, the primary relation $R_0$ does not have any foreign keys referencing other relations, and hence, Algorithm 3 would have constructed a single-relation model for $R_0$.

Algorithm 4 shows the pseudo-code of our data synthesis method for the case of no public relation in $\mathcal{R}$. It first utilizes the single-relation model constructed for $R_0$ to generate a synthetic version $R_0^*$ of $R_0$ (Line 1). After that, it inspects the foreign keys in $\mathcal{R}$ in descending order of $\prec$ (Lines 2-14). For each foreign key FK($R_i, R_j$) encountered, the algorithm differentiates two cases based on whether $R_i^*$ has been created previously when other foreign keys are processed (Line 3). If $R_i^*$ does not exist, then the algorithm initializes $R_i^* = \emptyset$, and retrieves the synthetic version $R_j^*$ of $R_j$ (which must have been generated due to the inspection order of the foreign keys). For each synthetic tuple $t^*$ in $R_j^*$, the algorithm samples a tuple group $G^*$ for $t^*$ using the graphical model for FK($R_i, R_j$), and inserts the tuples in $G^*$ into $R_i^*$ (Lines 3-7).

On the other hand, if $R_i^*$ already exists, then the processing of FK($R_i, R_j$) becomes more complicated. For example, consider the foreign keys shown in Figure 3, and assume that the public relations $P_1$ and $P_2$ do not exist. If we apply Algorithm 4 to process these foreign keys in descending order of $\prec$, then FK($R_3, R_1$) would be processed before FK($R_3, R_2$). Therefore, by the time that the algorithm examines FK($R_3, R_2$), a synthetic version $R_3^*$ of $R_3$ would have been constructed. In that case, if the algorithm handles FK($R_3, R_2$) as if $R_3^*$ does not exist, then it would generate another synthetic version of $R_3$. It is unclear how we can reconcile the differences between the two synthetic versions of $R_3$.

To address the above issue, when $R_i^*$ already exists, Algorithm 4 would sample synthetic tuple groups in a different manner as follows. First, it constructs a set $S = R_i^*$ (Line 9). Then, when it samples tuple groups from the model for FK($R_i, R_j$), it restricts the sample space to $S$, *i.e.*, only the tuples in $S$ has a chance to be sampled (Line 11). In addition, each time a tuple $t_i^*$ in $S$ is sampled, it is removed from $S$ to avoid being sampled again by the algorithm (Line 14). As such, the tuple groups generated for FK($R_i, R_j$) would consist of only the tuples in the previously generated $R_i^*$, and hence, we do not need to deal with two different synthetic versions of $R_i$.

## 5.3 Synthesis with Public Relations

Our solution in Section 5.2 only handles foreign keys between private relations. In what follows, we extend this solution to take into account foreign keys in the form of FK($R_i, P$), where $R_i$ and $P$ are private and public relations, respectively.

Observe that when Algorithm 3 processes FK($R_i, P$), it would (i) construct a graphical model for FK($R_i, P$), (ii) augment $P$ with estimated latent attributes pertinent to $R_i$, and (iii) create a DP single-relation model on the augmented $P$ (since $P$ has no private foreign key to other relations). If we are to generate synthetic data in relation to FK($R_i, P$), a naive approach is to first construct a synthetic version $P^*$ of $P$ based on the single-relation model for $P$, and then sample synthetic tuple groups from the graphical model for FK($R_i, P$). This, however, is unsatisfactory in terms of accuracy, since it does not exploit the fact that $P$ is a public relation and can be released directly. Another naive approach is to publish the augmented $P$, and then utilize the latent attributes therein to sample synthetic tuple groups for $R_i$ from the graphical model. Unfortunately, this violates DP, because the latent attributes of each tuple in $P$ convey private information and cannot be directly disclosed.

To address the limitations of the naive approaches above, we adopt an improved solution as follows. First, for each tuple $t$ in $P$, we use the DP single-relation model for $P$ to infer the latent attributes of $t$. This does not incur any privacy cost, since it uses only public information (*i.e., P*) and a DP model. Then, for each tuple $t$ with its inferred latent attributes, we generate a synthetic tuple group for $t$ by sampling from the graphical model for FK($R_i, P$). As such, we can avoid synthesizing $P$ while ensuring DP.

We incorporate this improved solution into Algorithm 4 as follows. First, we change the input of Algorithm 4 to allow $\mathcal{R}$ to have both private and public relations. Second, if the primary private relation $R_0$ has any foreign key to a public relation, we skip Line 1 of Algorithm 4, and leave the generation of $R_0^*$ to Lines 3-14. After that, when the algorithm processes private foreign keys in descending order of $\prec$, we handle each foreign key FK($R_i, R_j$) differently depending on whether $R_j$ is private or public. If $R_j$ is private, we process FK($R_i, R_j$) as in Lines 3-14 of Algorithm 4. Otherwise, we handle FK($R_i, R_j$) using the improved solution mentioned above, with one change: if a synthetic version $R_i^*$ of $R_i$ already exists, then we use $R_i^*$ as the sample space for the generation of tuple groups (see Lines 9-14 in Algorithm 4).

# 6 ENSURING DIFFERENTIAL PRIVACY

In this section, we explain how we can ensure $(\epsilon, \delta)$-DP in our solution by choosing appropriate parameters for each algorithm that we invoke. We start by introducing the concept of $L_2$ *sensitivity* and a known result on Gaussian-distributed noise.

*Definition 6.1 ($L_2$ Sensitivity [16]).* Let $f$ be a function that maps a database to a real vector. The $L_2$ *sensitivity* of $f$, denoted as $\Delta(f)$, is the maximum value of $\|f(\mathcal{R}) - f(\mathcal{R}')\|_2$ for any two neighboring database $\mathcal{R}$ and $\mathcal{R}'$, where $\|\cdot\|_2$ denotes the $L_2$ norm.

THEOREM 6.2 ([3]). *Let $\{f_1, \ldots, f_k\}$ be a set of functions. For any $i = 1, \ldots, k$, suppose that we inject independent Gaussian noise $\mathcal{N}(0, \sigma_i^2)$ into each element in $f_i$'s output. Then, the perturbed functions as a whole satisfy $(\epsilon, \delta)$-DP, if and only if*

$$\Phi\left(\frac{\gamma}{2} - \frac{\epsilon}{\gamma}\right) - e^{\epsilon} \cdot \Phi\left(-\frac{\gamma}{2} - \frac{\epsilon}{\gamma}\right) \leq \delta, \quad (6)$$

*where $\Phi$ is the cumulative distribution function of the standard normal distribution, and*

$$\gamma = \sqrt{\sum_{i=1}^{k} \left(\frac{\Delta(f_i)}{\sigma_i}\right)^2}.$$

Note that the left hand side of Equation (6) monotonically increases with $\gamma$. Therefore, if we are to inject Gaussian noise into a set of functions $\{f_1, \ldots, f_k\}$ to achieve $(\epsilon, \delta)$-DP, then by Theorem 6.2, we can set the noise scale $\sigma_i$ for each function $f_i$ as follows:

(1) Let $\gamma_{\max}$ be the largest $\gamma$ that satisfies Equation (6), for the given $\epsilon$ and $\delta$.
(2) Choose $\sigma_i$ to ensure that $\sum_{i=1}^{k} \left(\frac{\Delta(f_i)}{\sigma_i}\right)^2 \leq \gamma_{\max}^2$.

For convenience, we refer to $\sum_i \left(\frac{\Delta(f_i)}{\sigma_i}\right)^2$ as the *privacy consumption* of a function set $\{f_i\}$.

Observe that, whenever our algorithms access private information from $\mathcal{R}$, they always inject Gaussian noise into it for privacy protection. Therefore, if we can quantify the privacy consumption of each component of our solution, then we can set the parameters accordingly to ensure $(\epsilon, \delta)$-DP. In relation to this, we note that the privacy consumption of our algorithms depends on the private relations that they are applied on, because adding or removing one tuple in the primary private relation $R_0$ generally induces different amounts of changes in different secondary private relations. To take into account such dependencies, we define each relation's *tuple multiplier* and *group multiplier* as follows.

*Definition 6.3 (Tuple and Group Multipliers).* For any relation $R$ in $\mathcal{R}$, its *tuple multiplier* is the maximum number of tuples in $R$ that may change between two neighboring databases. In addition, its *group multiplier* with respect to a foreign key $\mathrm{FK}(R, R')$ is the maximum number of tuple groups induced by $\mathrm{FK}(R, R')$ that may change between two neighboring databases.

In what follows, we quantify the privacy consumption of our algorithms based on the tuple and group multipliers of the input relations. We starts with Algorithm 1. (Due to the space constraint, we include our proofs in Appendix C.)

Lemma 6.4. *Suppose that we apply Algorithm 1 on a relation $R$, considering the tuple groups induced by a foreign key* $\mathrm{FK}(R, R')$*. Let $\mu_g$ be the group multiplier of $R$ with respect to* $\mathrm{FK}(R, R')$*. Let $\tau$ be the maximum multiplicity of* $\mathrm{FK}(R, R')$*. Then, the privacy consumption of Algorithm 1 is:*

$$C_1(T, m_{\mathcal{M}}) = T \cdot \mu_g^2 \cdot \left( \frac{m_{\mathcal{M}} \cdot \tau^2}{\sigma_l^2} + \frac{1}{\sigma_{size}^2} + \frac{1}{\sigma_z^2} \right), \tag{7}$$

*where $T, \mathcal{M}, \sigma_z, \sigma_{size}, \sigma_l$ are the input parameters of Algorithm 1 and $m_{\mathcal{M}}$ is the number of latent marginals in $\mathcal{M}$.*

Next, we consider the DP single-relation modeling method [9] used in Line 1 of Algorithm 2 and Line 4 of Algorithm 3. Based on Lemma 2 in [9], we have the following result.

Corollary 6.5. *Suppose that we apply Algorithm 6 in [9] on a relation $R$ with a tuple multiplier $\mu_t$. Let $d$ be the number of attributes in $R$. Then, the privacy consumption of the algorithm is:*

$$C_{single}(d) = \mu_t^2 \cdot \left( \frac{1}{\sigma_{\mathcal{U}}^2} + \frac{2d(d-1)}{\sigma_R^2} + \frac{t \cdot k}{\sigma_h^2} + \frac{d+t}{\sigma_m^2} \right), \tag{8}$$

*where $\sigma_{\mathcal{U}}, \sigma_R, \sigma_h, \sigma_m, t, k$ are the parameters of the algorithm.*

Then, we have the following result for Algorithm 2.

Lemma 6.6. *Suppose that we apply Algorithm 2 on a relation $R$, considering the tuple groups induced by a foreign key* $\mathrm{FK}(R, R')$*. Let $\mu_t$ be the tuple multiplier of $R$, and $d$ be the number of attributes in $R$. Then, the privacy consumption of Algorithm 2 is:*

$$C_2(\mathrm{FK}(R, R')) = C_{single}(d) + C_1(T, 2d) + \sum_{i=1}^{T_C} C_1(1, 2d + i \cdot n_{inc}) + \mu_t^2 \cdot \frac{n_C \cdot T_C}{\sigma_{err}^2}. \tag{9}$$

Given Lemma 6.6 and Corollary 6.5, we can quantify the privacy consumption of Algorithm 3, by summing up the privacy consumption of (i) applying Algorithm 2 to process each private foreign key in $\mathcal{R}$ and (ii) applying the single-relation method on each relation that has no private foreign key but is referred to by at least one private relation. We can then set the parameters of PrivLava accordingly to ensure that PrivLava achieves $(\epsilon, \delta)$-DP with predefined $\epsilon$ and $\delta$. Note that once we obtain DP models from Algorithm 3, using the models to generate synthetic data does not incur any privacy cost. In addition, we can use the synthetic data to answer arbitrary queries without any privacy overhead.This is due to the *post-processing property* [17] of DP: if an algorithm satisfies $(\epsilon, \delta)$-DP, then adding any post-processing on the output of the algorithm does not degrade its privacy guarantee.

Lastly, we clarify how we divide the primary consumption of Algorithms 1 and 2 among their components. In Algorithm 1, we let the generation of noisy latent marginals (Lines 19-22) account for 80% of the primary consumption, and we split the remaining 20% between the generation of $p_{\text{size}}$ (Lines 13-18) and $p_{\text{size}}$ (Lines 8-12) in a ratio of 4 : 1. In other words, $\frac{m_{\mathcal{M}} \cdot \tau^2}{\sigma_l^2} : \frac{1}{\sigma_{size}^2} : \frac{1}{\sigma_z^2} = 20 : 4 : 1$. In addition, we set $T = 6$. In Algorithm 2, we allocate 20% of the privacy consumption to Line 1, 75% to the invocations of Algorithm 1, and the remaining 5% to Lines 9-14. We also set $T_C = 2$, $n_C = 400$, and $n_{\text{inc}} = d/4$. We determine these allocation ratios based on empirical calibrations across a range of different data sets, but these are not necessarily optimal. Identifying the optimal distribution of privacy budget is an interesting direction for future work.

# 7 EXPERIMENTS

## 7.1 Settings

**Data Sets.** We use two census databases from [10, 42] and the TPC-H benchmark [1] for experiments. The census databases contain census data collected in California and Île-de-France, respectively. Each of them contains a person relation $R_P$ and a household relation $R_H$, with a foreign key $FK(R_P, R_H)$ indicating the household that each person belongs to. We regard $R_H$ and $R_P$ as the primary and secondary private relations, respectively. Table 7 shows the statistics of the two census databases.

The TPC-H benchmark contains 8 relations and 8 foreign keys, as shown in Figure 4. We use Orders and Lineitem as the primary and secondary private relations, respectively, and regard the others as public relations. Note that the original TPC-H benchmark generates each foreign key based on a uniform distribution; as a result, it has minimum inter-relational correlations, which makes it unsuitable for evaluating the effectiveness of multi-relation synthesis. Motivated by this, we revise TPC-H to introduce inter-relational correlations as follows.

First, we assign a weight to each part-type and part-brand in Part. Then, for each tuple in Lineitem, we scale its price value by the product of the weights of its part-type and part-brand. In other words, we let the price of each lineitem depend on the type and brand of its part. In particular, the weight of each part-type (resp. part-brand) is chosen from $[0.2, 1]$ (resp. $[0.2, 2]$). Second, for each year $y$, we randomly select a 15% subset of the customers, and associate 40% of the orders in year $y$ to those customers; then, we distribute the remaining 60% orders randomly to the other 85% customers. This leads to correlations between customers and order-year. Third, for each year $y$, we drop each order in year $y$ with 75% probability, unless the order contains $y - 1992$ lineitems. This is to introduce correlations between the order sizes and order-year.

**Table 7: Summary of census datasets.**

**(a) California**

| Relation | # records | # attributes | domain size |
|---|---|---|---|
| Person | 1,690,642 | 23 | $\approx 6.77 \times 10^{12}$ |
| Household | 616,115 | 10 | $\approx 3.24 \times 10^{7}$ |

**(b) Île-de-France**

| Relation | # records | # attributes | domain size |
|---|---|---|---|
| Person | 4,297,133 | 14 | $\approx 1.84 \times 10^{10}$ |
| Household | 1,911,412 | 10 | $\approx 1.24 \times 10^{7}$ |

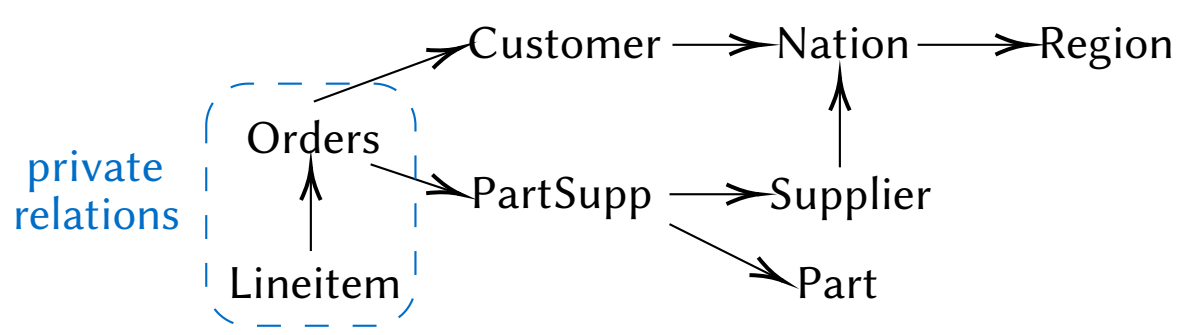

**Figure 4: The foreign key dependencies in TPC-H.**

With the above changes, we obtain a revised TPC-H benchmark with roughly $4 \times 10^6$ tuples in Lineitem and $1 \times 10^6$ tuples in Order.

Recall that our solution requires that the maximum multiplicity of each private foreign key $FK(R, R')$ is given. In practice, a data owner may choose an upper bound based on prior knowledge[2]; this assumption is adopted in previous work [46]. We follow the same assumption in our experiments. In particular, we examine the sizes of all tuple groups in $R$ with respect to $R'$, and calculate the 99% quantile of the group sizes. Then, we use this quantile as the maximum multiplicity of $FK(R, R')$.

**Baselines.** We consider three state-of-the-art solutions for DP data synthesis: PrivMRF [9], PrivBayes [49], and PB-PGM [34]. These three solutions are designed for single-relation synthesis, but we apply them on our data sets as follows. First, we use 80% of the privacy budget to generate a synthetic version of each private relation. Then, we spend the remaining 20% budget to generate, for each private foreign key $FK(R, R')$, a noisy histogram $H^*$ of the tuple group sizes in $R$ with respect to $FK(R, R')$. After that, we randomly associate the synthetic tuples for $R$ with the synthetic tuples for $R'$, while preserving the group size distribution in $H^*$.

In addition, we include a denormalization-based baseline, Denorm, which works as follows. It first uses 20% of the privacy budget to generate a noisy histogram of each foreign key's group size. Then, it joins all relevant relations in the input database into a denormalized relation, and generates a synthetic version $R^*$ of it using PrivMRF with the remaining 80% privacy budget. After that, it normalizes $R^*$ back into a set of relations. For example, suppose that $R^*$ is a synthetic version of $R_H \bowtie R_P$, where $R_H$ and $R_P$ are the household and person relation, respectively, in the California or Île-de-France data set. To normalize $R^*$, we first sort the tuples in $R^*$ in lexicographic order, with the household attributes preceding all person attributes. This results in a sorted sequence where adjacent tuples have similar household attributes. Then, we divide the sorted tuples into groups, following the distribution of group sizes observed in the noisy group size histogram. We regard each group $G$ as a household, and unify the household attributes in $G$ by first randomly choosing a tuple $t$ in $G$ and then copying $t$'s household attributes into all other tuples in $G$. The TPC-H database is handled in a similar manner.

**Privacy parameters.** In all of our experiments, we vary $\epsilon$ but fix $\delta = \frac{1}{n}$, where $n$ is the number of tuples in the secondary private relation. Similar settings are adopted in previous work [6, 9, 19, 22].

## 7.2 Experimental Results on Census Databases

Our first set of experiments focuses on the two census databases. We consider aggregate queries $Q(s, P_h, c, \{P_i\})$ in the following form.

Count the households $h$ that satisfy the following conditions:

- the size of $h$ equals $s$, and
- $h$ satisfies a predicate $P_h$ on its attributes, and
- $h$ contains $c$ persons $p_1, \ldots p_c$, such that each $p_i$ ($i = 1, \ldots, c$) satisfies a predicate $P_i$ on its attributes.

Intuitively, when $c = 1$, the accuracy of $Q$ on a synthetic database $\mathcal{R}^*$ indicates whether $\mathcal{R}^*$ preserves the correlations between the household and person relations. Meanwhile, when $c > 1$, $Q$ can also be used to evaluate whether $\mathcal{R}^*$ preserves the intra-group correlations between the household members.

We consider the cases of $c = 1$ and $c = 2$, with randomly selected $s$. For $P_h$ and $\{P_i\}$, we consider 1-attribute and 2-attribute conjuctive predicates, such that the condition on each attribute $A_j$ is $t[A_j] \in S_j$, where $t$ is a tuple and $S_j$ is a set of randomly selected values from $A_j$'s domain. We let $\frac{|S_j|}{|A_j|} = (0.2)^{1/k}$, where $k$ is the total number of conditions in $P_h$ and $\{P_i\}$. In other words, we set the size of each $S_j$ to a constant, such that the selectivity of all predicates combined is 0.2 on uniformly distributed data. We measure the accuracy of each query $Q$ by its *relative error*, which is defined as

$$\frac{\text{absolute error of } Q}{\max\{\text{actual result of } Q,\ 0.01 \cdot \text{total number of households}\}},$$

where "0.01 · total number of households" is a regularization term to mitigate the effects of excessively small query results.

Figure 5 (resp. Figure 6) shows the errors of each method averaged over 10,000 queries on California (resp. Île-de-France). Observe that, regardless of whether $c = 1$ or $c = 2$ and whether each predicate involves 1 or 2 attributes, PrivLava significantly outperforms its competitors in terms of query accuracy. This demonstrates the effectiveness of PrivLava in modeling the inter-relational and intra-group correlations in the input data. The accuracy of PrivLava improves when the privacy budget $\epsilon$ increases, since a larger $\epsilon$ allows PrivLava to use smaller noise for data synthesis. In contrast, the accuracy of the baseline methods shows little improvement with the increase of $\epsilon$, which indicates that their inability to model foreign keys is the main cause of their query errors. In addition, Denorm outperforms other baselines, but is still inferior to PrivLava, which shows that the denormalization approach of Denorm helps reserve inter-relational correlations but is not as effective as PrivLava's latent variable approach.

[2]When such prior knowledge is unavailable, the data owner may use a differentially private algorithm to decide a suitable upper bound, e.g., by computing an approximate 99% quantile of the degrees of the foreign key; previous work [15, 25] has presented algorithms for this purpose.

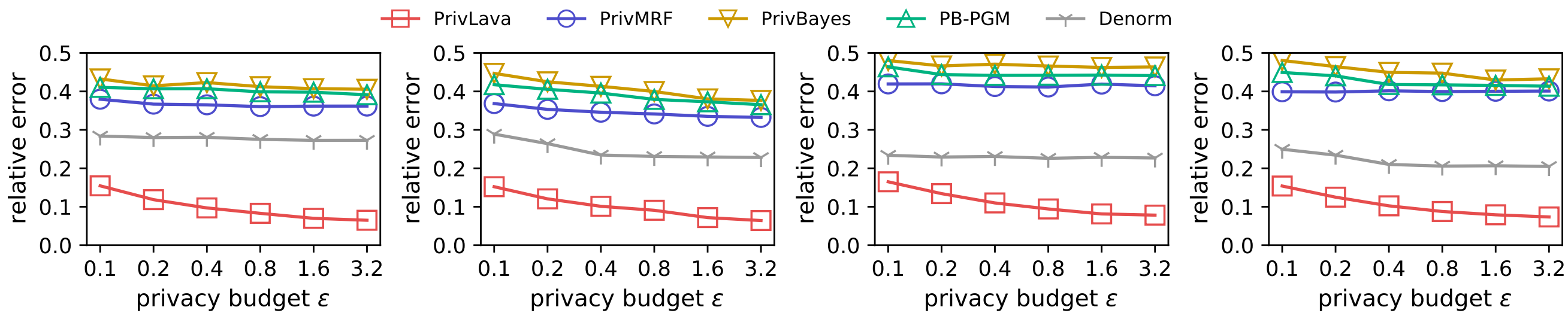

(a) $c = 1$, 1-attribute predicates (b) $c = 1$, 2-attribute predicates (c) $c = 2$, 1-attribute predicates (d) $c = 2$, 2-attribute predicates

**Figure 5: California: relative error vs. $\epsilon$.**

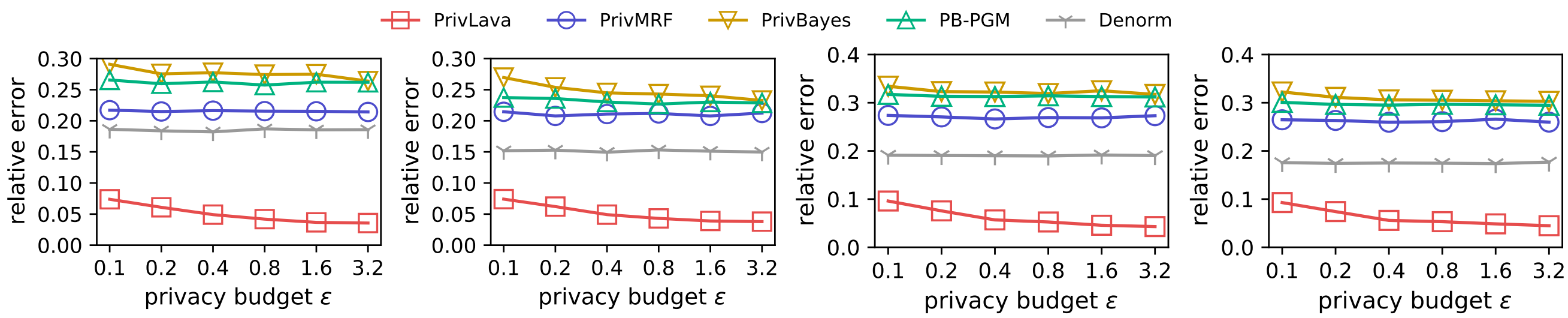

(a) $c = 1$, 1-attribute predicates (b) $c = 1$, 2-attribute predicates (c) $c = 2$, 1-attribute predicates (d) $c = 2$, 2-attribute predicates

**Figure 6: Île-de-France: relative error vs. $\epsilon$.**

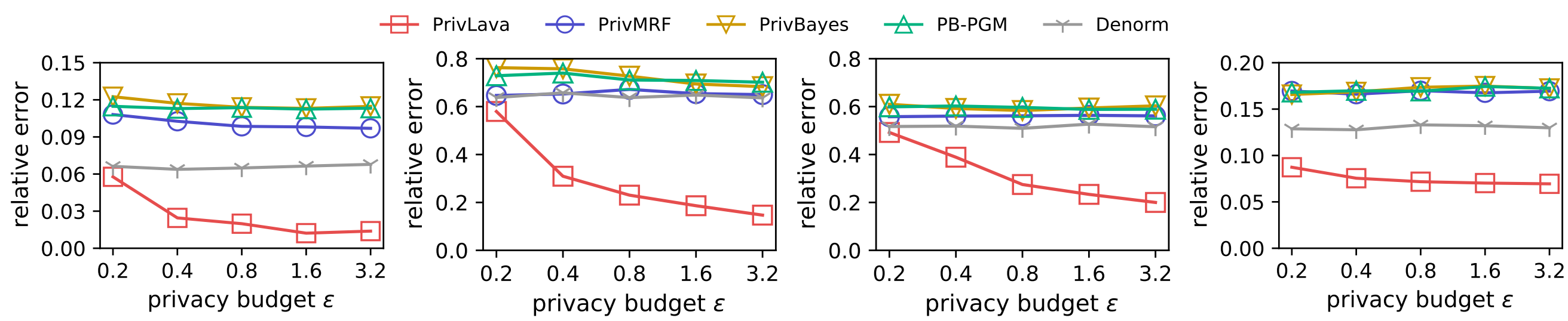

(a) Q4 (b) Q5 (c) Q7 (d) Q9

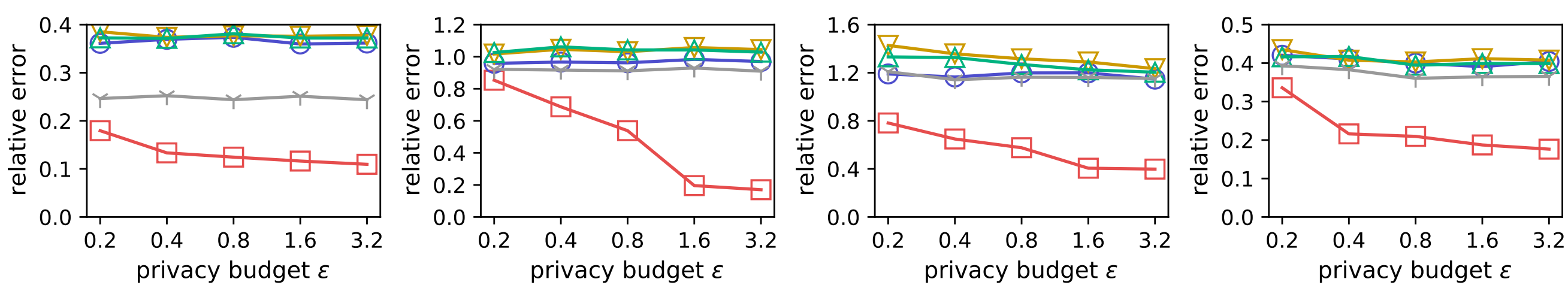

(e) Q12 (f) Q14 (g) Q17 (h) Q19

**Figure 7: TPC-H: relative error vs. $\epsilon$.**

## 7.3 Experimental Results on TPC-H

In our second set of experiments, we evaluate each method on 8 aggregate queries from the TPC-H benchmark. Each query $Q$ returns either a value or a set of values generated with GROUP BY. We measure the accuracy of $Q$ by the average relative error of the value(s) that it returns. In addition, we remove the regularization term from the definition of relative error, since all values from TPC-H queries are sufficiently large.

Figure 7 shows the accuracy of each method as a function of $\epsilon$, averaged over 10 runs. Observe that PrivLava consistently outperforms its competitors, in most cases by a large margin. This is consistent with the results in Figure 5 and Figure 6. Interestingly, there are a few queries (*e.g.*, Q7, Q14, Q17) for which PrivLava's accuracy degrades noticeably when $\epsilon$ is small. The main reason is that some components of those queries have small selectivity, and hence, they can be answered accurately only if our modeling of the input data is highly accurate, which is difficult under a small privacy budget $\epsilon$. For example, the result of Q17 contains one value that is computed over merely 0.01% of the tuples in Lineitem. Nonetheless, even with a modest $\epsilon = 1.6$, PrivLava is already able to substantially outperform its competitors on all queries.

Denorm shows noticeable improvement over other baselines in Q4, Q9, and Q12. For all other queries, however, Denorm only offers comparable query accuracy to other baselines. This is consistent with our observation from the experimental results in Section 7.2: Denorm is able to capture some inter-relational correlations, but it is inferior to PrivLava in terms of modeling accuracy.

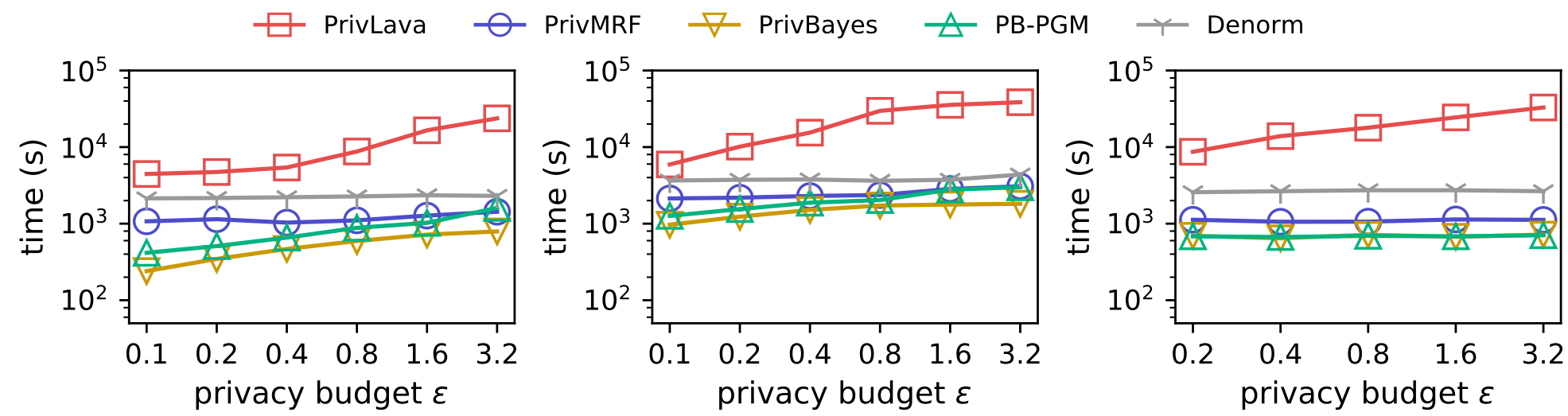

(a) California (b) Île-de-France (c) TPC-H

**Figure 8: Computation cost vs. $\epsilon$.**

### 7.4 Computation Cost

Our last set of experiments evaluates the computation time of each method on our three data sets, using a Linux machine with a 32-core 2.9GHz CPU and an Nvidia A10 GPU. All methods are implemented using Python. Figure 8 shows the processing cost of each method, averaged over 10 runs. The computation cost of PrivLava increases with the privacy budget $\epsilon$, since (i) a larger $\epsilon$ allows PrivLava to use larger marginals for modeling (see Equation 5), and (ii) such marginals are more time-consuming to process. PrivLava incurs a higher computation time than other baselines, but it is still able to finish processing the input data set within 12 hours in all cases. PrivLava's computation time is acceptable, since data synthesis only incurs a one-time cost, and the synthetic data generated can be used repeatedly by downstream applications.

## 8 RELATED WORK

As mentioned in Section 1, there exist a sizable number of methods for synthesizing data under DP [7, 9, 13, 21, 23, 29, 34, 36, 39, 47–50], but all of them focus on the single-relation cases. Among these methods, the most comparable to our proposed approach is PrivMRF [9], which also models the input data using a similar family of graphical models. However, PrivMRF differs significantly from PrivLava in the sense that the graphical models do not include any latent variables, whereas the use of latent variables in PrivLava is the key ingredient in its modeling of foreign key constraints. The usage of latent variables also makes the model construction algorithm of PrivLava considerably more complex than those in [9]. Existing work [13, 21, 23, 29, 36, 47] has also proposed DP single-relation synthesis techniques using other models (*e.g.*, Copula functions [29], generative adversarial networks [23, 39, 48]), but the performance of these techniques is generally not as good as methods based on graphical models, as shown in [9].

In the context of non-DP data synthesis, existing work [19, 20, 24, 26, 31, 32] has presented a number of solutions for handling foreign keys. Specifically, the technique in [20] assumes that there are hard constraints on the non-key attributes in each table, and utilizes those constraints to generate foreign keys between synthetic relations. In practice, however, it is unclear where we obtain such constraints in a privacy preserving manner, and whether they are sufficient to preserve the correlations induced by foreign keys. Meanwhile, previous work [24, 26, 31, 32] propose to retain the foreign keys in the input data, and perturb the non-key attributes for privacy protection. Unfortunately, these approaches fail to provide strong privacy guarantees.

In addition, there is a long line of research [5, 6, 8, 12, 15, 22, 25, 33, 35, 37, 40, 41, 45, 46] on constructing database engines under DP using output perturbation. As we explain in Section 1, however, such database engines may only answer a limited number of queries accurately, since each query inevitably reveals some private information about the underlying data. Note that this limitation is inherent to output perturbation: Dinur and Nissim [14] has proved that all output perturbation methods would suffer from the same issue under any reasonably strong privacy notion.

Finally, previous work [30] proposed an algorithm for generating database benchmarks under DP. In particular, given a database $\mathcal{D}$ and a workload $Q$ of queries, the algorithm produces another database $\mathcal{D}'$ under DP, such that if we use a database engine to process $Q$ on $\mathcal{D}$ and $\mathcal{D}'$, respectively, each query would have similar processing time on both databases. In other words, the algorithm aims to preserve the costs of queries in $Q$, instead of the accuracy of the query results. Therefore, the solution in [30] is inapplicable in our setting.

## 9 CONCLUSION

This paper presents PrivLava, the first solution for synthesizing relational data with foreign keys under differential privacy (DP). PrivLava models the input database using graphical models, with latent variables included to capture the complex correlations induced by foreign key constraints. It supports databases with both private and public relations, with any foreign keys that form a directed acyclic graph. Extensive experiments on real data show that PrivLava offers significantly higher query accuracy on synthetic data than competing methods do.

The main limitation of PrivLava is that it requires a substantial amount of data to train accurate DP models of foreign keys. In particular, it requires generating a relatively large number of observed and latent marginals from the input database, and it injects Gaussian noise into each marginal to achieve DP. When the number of tuples in the input database is small, the noise injected into each marginal may overwhelm the signal therein (especially in latent marginals); in that case, PrivLava is unable to model foreign keys effectively. In future work, we plan to investigate how this issue could be mitigated, and to extend PrivLava to take into account other types of constraints on relational data, *e.g.*, denial constraints [11].

**Acknowledgements.** This research/project is supported by the National Research Foundation, Singapore under its Strategic Capability Research Centres Funding Initiative. Any opinions, findings and conclusions or recommendations expressed in this material are those of the author(s) and do not reflect the views of National Research Foundation, Singapore. This work is also supported in part by the UKRI Prosperity Partnership Scheme (FAIR) under the EPSRC Grant EP/V056883/1, and the Alan Turing Institute.

## REFERENCES

[1] 2022. TPC-H benchmark homepage. https://www.tpc.org/tpch/.

[2] Gergely Acs, Claude Castelluccia, and Rui Chen. 2012. Differentially private histogram publishing through lossy compression. In *2012 IEEE 12th International Conference on Data Mining*. IEEE, 1–10.

[3] Borja Balle and Yu-Xiang Wang. 2018. Improving the gaussian mechanism for differential privacy: Analytical calibration and optimal denoising. In *International Conference on Machine Learning*. PMLR, 394–403.

[4] Johes Bater, Gregory Elliott, Craig Eggen, Satyender Goel, Abel N Kho, and Jennie Rogers. 2017. SMCQL: Secure Query Processing for Private Data Networks. *Proc. VLDB Endow.* 10, 6 (2017), 673–684.

[5] Johes Bater, Xi He, William Ehrich, Ashwin Machanavajjhala, and Jennie Rogers. 2018. Shrinkwrap: efficient sql query processing in differentially private data federations. *Proceedings of the VLDB Endowment* 12, 3 (2018).

[6] Johes Bater, Yongjoo Park, Xi He, Xiao Wang, and Jennie Rogers. 2020. Saqe: practical privacy-preserving approximate query processing for data federations. *Proceedings of the VLDB Endowment* 13, 12 (2020), 2691–2705.

[7] Vincent Bindschaedler, Reza Shokri, and Carl A Gunter. 2017. Plausible Deniability for Privacy-Preserving Data Synthesis. *Proceedings of the VLDB Endowment* 10, 5 (2017).

[8] Jaroslaw Błasiok, Mark Bun, Aleksandar Nikolov, and Thomas Steinke. 2019. Towards instance-optimal private query release. In *Proceedings of the Thirtieth Annual ACM-SIAM Symposium on Discrete Algorithms*. SIAM, 2480–2497.

[9] Kuntai Cai, Xiaoyu Lei, Jianxin Wei, and Xiaokui Xiao. 2021. Data synthesis via differentially private markov random fields. *Proceedings of the VLDB Endowment* 14, 11 (2021), 2190–2202.

[10] Minnesota Population Center. 2020. Integrated Public Use Microdata Series, International: Version 7.3 [dataset]. Minneapolis, MN: IPUMS. https://doi.org/10.18128/D020.V7.3.

[11] Jan Chomicki and Jerzy Marcinkowski. 2005. Minimal-change integrity maintenance using tuple deletions. *Information and Computation* 197, 1-2 (2005), 90–121.

[12] Graham Cormode, Tejas Kulkarni, and Divesh Srivastava. 2018. Marginal release under local differential privacy. In *Proceedings of the 2018 International Conference on Management of Data*. 131–146.

[13] Graham Cormode, Cecilia Procopiuc, Divesh Srivastava, and Thanh TL Tran. 2012. Differentially private summaries for sparse data. In *Proceedings of the 15th International Conference on Database Theory*. 299–311.

[14] Irit Dinur and Kobbi Nissim. 2003. Revealing Information While Preserving Privacy. In *Proceedings of the Twenty-Second ACM SIGMOD-SIGACT-SIGART Symposium on Principles of Database Systems*. 202–210. https://doi.org/10.1145/773153.773173

[15] Wei Dong, Juanru Fang, Ke Yi, Yuchao Tao, and Ashwin Machanavajjhala. 2022. R2T: Instance-optimal Truncation for Differentially Private Query Evaluation with Foreign Keys. In *Proc. ACM SIGMOD International Conference on Management of Data*.

[16] Cynthia Dwork, Frank McSherry, Kobbi Nissim, and Adam Smith. 2006. Calibrating noise to sensitivity in private data analysis. In *Theory of cryptography conference*. Springer, 265–284.

[17] Cynthia Dwork, Aaron Roth, et al. 2014. The algorithmic foundations of differential privacy. *Foundations and Trends® in Theoretical Computer Science* 9, 3–4 (2014), 211–407.

[18] Marco Gaboardi, Emilio Jesús Gallego Arias, Justin Hsu, Aaron Roth, and Zhiwei Steven Wu. 2014. Dual query: Practical private query release for high dimensional data. In *International Conference on Machine Learning*. PMLR, 1170–1178.

[19] Chang Ge, Shubhankar Mohapatra, Xi He, and Ihab F Ilyas. 2021. Kamino: constraint-aware differentially private data synthesis. *Proceedings of the VLDB Endowment* 14, 10 (2021), 1886–1899.

[20] Amir Gilad, Shweta Patwa, and Ashwin Machanavajjhala. 2021. Synthesizing linked data under cardinality and integrity constraints. In *Proceedings of the 2021 International Conference on Management of Data*. 619–631.

[21] Moritz Hardt, Katrina Ligett, and Frank McSherry. 2012. A simple and practical algorithm for differentially private data release. *Advances in neural information processing systems* 25 (2012).

[22] Noah Johnson, Joseph P Near, and Dawn Song. 2018. Towards practical differential privacy for SQL queries. *Proceedings of the VLDB Endowment* 11, 5 (2018), 526–539.

[23] James Jordon, Jinsung Yoon, and Mihaela Van Der Schaar. 2018. PATE-GAN: Generating synthetic data with differential privacy guarantees. In *International conference on learning representations*.

[24] Christopher T Kenny, Shiro Kuriwaki, Cory McCartan, Evan TR Rosenman, Tyler Simko, and Kosuke Imai. 2021. The use of differential privacy for census data and its impact on redistricting: The case of the 2020 US Census. *Science advances* 7, 41 (2021), eabk3283.

[25] Ios Kotsogiannis, Yuchao Tao, Xi He, Maryam Fanaeepour, Ashwin Machanavajjhala, Michael Hay, and Gerome Miklau. 2019. Privatesql: a differentially private sql query engine. *Proceedings of the VLDB Endowment* 12, 11 (2019), 1371–1384.

[26] Amy Lauger, Billy Wisniewski, and Laura McKenna. 2014. Disclosure avoidance techniques at the US Census Bureau: Current practices and research. *Center for Disclosure Avoidance Research, US Census Bureau* (2014).

[27] Chao Li, Michael Hay, Gerome Miklau, and Yue Wang. 2014. A Data-and Workload-Aware Algorithm for Range Queries Under Differential Privacy. *Proceedings of the VLDB Endowment* 7, 5 (2014).

[28] Chao Li, Gerome Miklau, Michael Hay, Andrew McGregor, and Vibhor Rastogi. 2015. The matrix mechanism: optimizing linear counting queries under differential privacy. *The VLDB journal* 24, 6 (2015), 757–781.

[29] Haoran Li, Li Xiong, and Xiaoqian Jiang. 2014. Differentially private synthesization of multi-dimensional data using copula functions. In *Advances in database technology: proceedings. International conference on extending database technology*, Vol. 2014. NIH Public Access, 475.

[30] Wentian Lu, Gerome Miklau, and Vani Gupta. 2014. Generating private synthetic databases for untrusted system evaluation. In *2014 IEEE 30th International Conference on Data Engineering*. IEEE, 652–663.

[31] Paul M Massell and Jeremy M Funk. 2007. *Recent Developments in the Use of Noise for Protecting Magnitude Data Tables: Balancing to Improve Data Quality and Rounding that Preserves Rrotection.* US Census Bureau [custodian]].

[32] Laura McKenna. 2018. *Disclosure Avoidance Techniques Used for the 1970 through 2010 Decennial Censuses of Population and Housing.* Technical Report. US Census Bureau, Center for Economic Studies.

[33] Ryan McKenna, Gerome Miklau, Michael Hay, and Ashwin Machanavajjhala. 2018. Optimizing error of high-dimensional statistical queries under differential privacy. *Proceedings of the VLDB Endowment* 11, 10 (2018), 1206–1219.

[34] Ryan McKenna, Daniel Sheldon, and Gerome Miklau. 2019. Graphical-model based estimation and inference for differential privacy. In *International Conference on Machine Learning*. PMLR, 4435–4444.

[35] Frank D McSherry. 2009. Privacy integrated queries: an extensible platform for privacy-preserving data analysis. In *Proceedings of the 2009 ACM SIGMOD International Conference on Management of data*. 19–30.

[36] Noman Mohammed, Rui Chen, Benjamin CM Fung, and Philip S Yu. 2011. Differentially private data release for data mining. In *Proceedings of the 17th ACM SIGKDD international conference on Knowledge discovery and data mining*. 493–501.

[37] Prashanth Mohan, Abhradeep Thakurta, Elaine Shi, Dawn Song, and David Culler. 2012. GUPT: privacy preserving data analysis made easy. In *Proceedings of the 2012 ACM SIGMOD International Conference on Management of Data*. 349–360.

[38] Aleksandar Nikolov, Kunal Talwar, and Li Zhang. 2016. The geometry of differential privacy: The small database and approximate cases. *SIAM J. Comput.* 45, 2 (2016), 575–616.

[39] Noseong Park, Mahmoud Mohammadi, Kshitij Gorde, Sushil Jajodia, Hongkyu Park, and Youngmin Kim. 2018. Data Synthesis based on Generative Adversarial Networks. *Proceedings of the VLDB Endowment* 11, 10 (2018).

[40] Davide Proserpio, Sharon Goldberg, and Frank McSherry. 2014. Calibrating data to sensitivity in private data analysis: A platform for differentially-private analysis of weighted datasets. *Proceedings of the VLDB Endowment* 7, 8 (2014), 637–648.

[41] Wahbeh Qardaji, Weining Yang, and Ninghui Li. 2014. Priview: practical differentially private release of marginal contingency tables. In *Proceedings of the 2014 ACM SIGMOD international conference on Management of data*. 1435–1446.

[42] Steven Ruggles, Sarah Flood, Ronald Goeken, Megan Schouweiler, and Matthew Sobek. 2022. IPUMS USA: VERSION 12.0 [dataset]. Minneapolis, MN: IPUMS. https://doi.org/10.18128/D010.V12.0.

[43] Yuchao Tao, Xi He, Ashwin Machanavajjhala, and Sudeepa Roy. 2020. Computing local sensitivities of counting queries with joins. In *Proceedings of the 2020 ACM SIGMOD International Conference on Management of Data*. 479–494.

[44] Martin J Wainwright, Michael I Jordan, et al. 2008. Graphical models, exponential families, and variational inference. *Foundations and Trends® in Machine Learning* 1, 1–2 (2008), 1–305.

[45] Tianhao Wang, Bolin Ding, Jingren Zhou, Cheng Hong, Zhicong Huang, Ninghui Li, and Somesh Jha. 2019. Answering multi-dimensional analytical queries under

local differential privacy. In *Proceedings of the 2019 International Conference on Management of Data*. 159–176.
[46] Royce J Wilson, Celia Yuxin Zhang, William Lam, Damien Desfontaines, Daniel Simmons-Marengo, and Bryant Gipson. 2020. Differentially Private SQL with Bounded User Contribution. *Proceedings on Privacy Enhancing Technologies* 2 (2020), 230–250.
[47] Chugui Xu, Ju Ren, Yaoxue Zhang, Zhan Qin, and Kui Ren. 2017. DPPro: Differentially private high-dimensional data release via random projection. *IEEE Transactions on Information Forensics and Security* 12, 12 (2017), 3081–3093.
[48] Lei Xu, Maria Skoularidou, Alfredo Cuesta-Infante, and Kalyan Veeramachaneni. 2019. Modeling tabular data using conditional gan. *Advances in Neural Information Processing Systems* 32 (2019).
[49] Jun Zhang, Graham Cormode, Cecilia M Procopiuc, Divesh Srivastava, and Xiaokui Xiao. 2017. Privbayes: Private data release via bayesian networks. *ACM Transactions on Database Systems (TODS)* 42, 4 (2017), 1–41.
[50] Jun Zhang, Xiaokui Xiao, and Xing Xie. 2016. Privtree: A differentially private algorithm for hierarchical decompositions. In *Proceedings of the 2016 International Conference on Management of Data*. 155–170.

# A PARAMETER ESTIMATION

In this section, we elaborate our parameter estimation method for $\theta$ in Line 23 of Algorithm 1. Recall that our objective of parameter estimation is to maximize the overall likelihood $\sum_G \log p(G)$, where $p(G)$ is defined in Eq. (2). Towards this end, We use EM, which attempts to maximize the following expected log likelihood in each iteration $j \in [T]$:

$$Q^{(j)} = \sum_{G \in R} \sum_{z} p^{(j)}(z \mid G) \log p(G, z), \tag{10}$$

where $p^{(j)}(z \mid G)$ is the conditional latent variable distribution $p(z \mid G)$ in the $j$-th iteration, as defined in Eq. (3), and $p(G, z)$ is as defined in Eq. (4).

Let $p(t, z)$ be as defined in Eq. (1) and $\theta$ be the parameter of $p$. Let $p_{\text{tuple}}(t, z; \theta)$ denote $p(t, z)$ when it is parameterized with $\theta$, and $p_{\text{tuple}}(t \mid z; \theta) = \frac{p_{\text{tuple}}(t,z;\theta)}{\sum_t p_{\text{tuple}}(t,z;\theta)}$. For convenience, we omit $\theta$ in $p_{\text{tuple}}(\cdot)$ when the context is clear. We will show that the estimation results in the $j$-th iteration of Algorithm 1 are:

$$p_Z^{(j)}, p_{\text{size}}^{(j)}, p_{\text{tuple}}^{(j)} = \underset{p_Z, p_{\text{size}}, p_{\text{tuple}}}{\arg\max}\ Q^{(j)}. \tag{11}$$

As mentioned in Section 4.2, Line 23 of Algorithm 1 invokes the gradient descent method in [34], which is referred to as PGM. We first briefly explain PGM, and then prove that the invocation of PGM in Line 23 of Algorithm 1 returns the parameters $\theta^{(j)}$ of $p_{\text{tuple}}^{(j)}$. Given a collection $\mathcal{M}$ of marginals, PGM identifies the parameters $\theta$ of $p_{\text{tuple}}$ by minimizing $L = \|\mathcal{M}_R - \mathcal{M}_\theta\|_2$, where $\mathcal{M}_R$ is the noisy counts of the relation $R$ and $\mathcal{M}_\theta$ is the marginal counts of $p_{\text{tuple}}$. In what follows, we first transform the problem of maximizing Eq. (10) into the problem of maximizing a new objective function regarding $p_{\text{tuple}}(t, z)$, as shwon in Lemma A.1. Then, we show in Theorem A.2 that the new objective is concave, and thus, can be maximized with gradient ascent methods such as PGM.

With Eq. (10), (11), and (4), we have:

$$\begin{aligned}
&p_Z^{(j)}, p_{\text{size}}^{(j)}, p_{\text{tuple}}^{(j)} \\
&= \underset{p_Z, p_{\text{size}}, p_{\text{tuple}}}{\arg\max} \sum_{G \in R} \sum_{z} p^{(j)}(z \mid G) \log p(G, z) \\
&= \underset{p_Z, p_{\text{size}}, p_{\text{tuple}}}{\arg\max} \sum_{G \in R} \sum_{z} p^{(j)}(z \mid G) \\
&\qquad \cdot \log \left( p_Z(z) \cdot p_{\text{size}}(|G| \mid z) \cdot \prod_{t \in G} p(t \mid z) \right) \\
&= \underset{p_Z, p_{\text{size}}, p_{\text{tuple}}}{\arg\max} \sum_{G \in R} \sum_{z} p^{(j)}(z \mid G) \Bigg( \log p_Z(z) + \log p_{\text{size}}(|G| \mid z) \\
&\qquad + \sum_{t \in G} \log p_{\text{tuple}}(t \mid z; \theta) \Bigg).
\end{aligned}$$

Notice that we may compute $p_Z^{(j)}, p_{\text{size}}^{(j)}, p_{\text{tuple}}^{(j)}$ separately as follows:

$$\begin{aligned}
p_Z^{(j)} &= \underset{p_Z}{\arg\max} \sum_{G \in R} \sum_{z} p^{(j)}(z \mid G) \log p_Z(z), \\
p_{\text{size}}^{(j)} &= \underset{p_{\text{size}}}{\arg\max} \sum_{G \in R} \sum_{z} p^{(j)}(z \mid G) \log p_{\text{size}}(|G| \mid z), \\
p_{\text{tuple}}^{(j)} &= \underset{p_{\text{tuple}}}{\arg\max} \sum_{G \in R} \sum_{z} \sum_{t \in G} p^{(j)}(z \mid G) \log p_{\text{tuple}}(t \mid z; \theta).
\end{aligned} \tag{12}$$

Let $N$ denote the number of groups in $R$. By Gibbs' inequality, we have:

$$\begin{aligned}
p_Z^{(j)}(z) &= \frac{1}{N} \sum_{G \in R} p^{(j)}(z \mid G), \\
p_{\text{size}}^{(j)}(s \mid z) &= \frac{\sum_{G \in R} p^{(j)}(z \mid G) \mathbb{I}(|G| = s)}{\sum_{G \in R} p^{(j)}(z \mid G)}.
\end{aligned}$$

We first solve Eq. (12). By Eq. (1), we can rewrite $p_{\text{tuple}}(t, z; \theta)$ as:

$$p_{\text{tuple}}(t, z; \theta) = \frac{1}{A(\theta)} \prod_{c \in M \in \mathcal{M} \wedge c \leftrightarrow (t,z)} \exp(\theta[c]), \tag{13}$$

where $A(\theta)$ is a constant such that $\sum_{t,z} p_{\text{tuple}}(t, z; \theta) = 1$. Similarly, $p_{\text{tuple}}(t \mid z; \theta)$ can be rewritten as:

$$p_{\text{tuple}}(t \mid z; \theta) = \frac{1}{A'(\theta, z)} \prod_{c \in M \in \mathcal{M} \wedge c \leftrightarrow (t,z)} \exp(\theta[c]), \tag{14}$$

where $A'(\theta, z)$ is also a constant such that $\sum_t p_{\text{tuple}}(t \mid z; \theta) = 1$ for any $z$.

Observe that $p_{\text{tuple}}$ is determined by $\theta$ when the structure $\mathcal{M}$ is fixed. Therefore, Eq. (12) is equivalent to:

$$\theta^{(j)} = \underset{\theta}{\arg\max} \sum_{G \in R} \sum_{z} \sum_{t \in G} p^{(j)}(z \mid G) \log p_{\text{tuple}}(t \mid z; \theta). \tag{15}$$

To solve this equation, we have the following lemma. (The proof is given in Appendix B.)

Lemma A.1. *Let $M_Z$ denote the collection of latent variables. Given some latent variable distribution $p(z|G)$ for each group and a marginal set $\mathcal{M}$ such that there exist $M \in \mathcal{M}$ and $M_Z \subseteq M$, let:*

$$L_1(\theta) = \sum_{G\in R}\sum_{z}\sum_{t\in G} p(z \mid G) \log p_{tuple}(t \mid z; \theta), \tag{16}$$

$$L_2(\theta) = \sum_{G\in R}\sum_{z}\sum_{t\in G} p(z \mid G) \log p_{tuple}(t, z; \theta), \tag{17}$$

$$\theta^* = \arg\max_{\theta} L_2. \tag{18}$$

*Then, $\theta^*$ is a solution of $\arg\max_\theta L_1$.*

In the case of a single latent variable, $M_Z$ is a subset of any latent marginals. In the case of multiple latent variables, we require that $\mathcal{M}$ contains $M_Z$ in Section 4.6. Therefore, we may apply this lemma to solve Eq. (15). Let:

$$L^{(j)}(\theta) = \sum_{G\in R}\sum_{z}\sum_{t\in G} p^{(j)}(z \mid G) \log p_{\text{tuple}}(t, z; \theta).$$

We have:

$$\begin{aligned}\theta^{(j)} &= \arg\max_{\theta} L^{(j)}(\theta) \\ &= \arg\max_{\theta} \sum_{G\in R}\sum_{z}\sum_{t\in G} p^{(j)}(z \mid G) \log p_{\text{tuple}}(t, z; \theta).\end{aligned} \tag{19}$$

This turns out to be a convex optimization problem, as shown in the following theorem.

Theorem A.2. *Given some latent variable distribution $p(z \mid G)$ of each group $G$, let:*

$$L(\theta) = \sum_{G\in R}\sum_{z}\sum_{t\in G} p(z \mid G) \log p_{tuple}(t, z; \theta). \tag{20}$$

*Then, $L$ is concave and its partial derivative of each count $c$ is:*

$$\frac{\partial L}{\partial\theta[c]} = \sum_{G\in R}\sum_{t\in G, z: c\leftrightarrow(t,z)} p^{(j)}(z \mid G) - n \sum_{t,z: c\leftrightarrow(t,z)} p_{tuple}(t, z; \theta). \tag{21}$$

Proof. By Eq. (13) and (20),

$$\begin{aligned}L(\theta) &= \sum_{G\in R}\sum_{z}\sum_{t\in G} p^{(j)}(z \mid G)\left(\sum_{c\in M\in\mathcal{M}\wedge c\leftrightarrow(t,z)} \theta[c] - \log A(\theta)\right) \\ &= \sum_{G\in R}\sum_{z}\sum_{t\in G} p^{(j)}(z \mid G) \sum_{c\in M\in\mathcal{M}\wedge c\leftrightarrow(t,z)} \theta[c] - n\log A(\theta),\end{aligned}$$

where $n$ is the number of tuples in $R$. The first term is a linear function since $p^{(j)}(z \mid G)$ are all fixed values. By linearity, it is also concave. Besides, it is well known that $\log A(\theta)$ is convex w.r.t $\theta$ in the literature of graphical models. Therefore, $L(\theta)$ is concave. For each count $c$, the partial derivative of $\theta[c]$ is:

$$\frac{\partial L}{\partial\theta[c]} = \sum_{G\in R}\sum_{t\in G, z: c\leftrightarrow(t,z)} p^{(j)}(z \mid G) - n \sum_{t,z: c\leftrightarrow(t,z)} p_{\text{tuple}}(t, z; \theta),$$

where the first term is due to the linearity and the second term is given by the property $\partial \log A(\theta)/\partial\theta[c] = \sum_{t,z: c\leftrightarrow(t,z)} p_{\text{tuple}}(t, z; \theta)$. □

Theorem A.2 shows that Eq. (19) is a convex optimization problem and we may apply any gradient ascent method to estimate $\theta^{(j)}$. Particularly, the first term of the derivative Eq. (21) is the value of count $c$ of the relation and the second term is a marginal count of $p_{\text{tuple}}$. We may calculate the second term with any marginal inference algorithm (e.g. junction tree algorithm).

Moreover, since the derivative is the difference between the data marginal counts and the model marginal counts, our gradient method is the same as Algorithm 1 of PGM [34] by extending the concept of marginals to include latent marginals. This explains Line 23 of Algorithm 1.

## B PROOF OF LEMMA A.1

By Eq. (13) and (17),

$$\begin{aligned}L_2(\theta) &= \sum_{G\in R}\sum_{z}\sum_{t\in G} p(z \mid G)\Bigg(\sum_{c\in M\in\mathcal{M}\wedge c\leftrightarrow(t,z)} \theta[c] \\ &\qquad - \log A(\theta)\Bigg) \\ &= \sum_{G\in R}\sum_{z}\sum_{t\in G} p(z \mid G) \sum_{c\in M\in\mathcal{M}\wedge c\leftrightarrow(t,z)} \theta[c] \\ &\qquad - n\log A(\theta),\end{aligned}$$

where $n$ is the number of tuples in $R$. The first term is a linear function since $p(z \mid G)$ are all fixed values. By linearity, it is also concave. Besides, it is well known that $\log A(\theta)$ is convex w.r.t $\theta$ in the literature of graphical models. Therefore, $L_2$ is concave and the partial derivative $\partial L_2/\partial\theta[c]$ are all zero at $\theta^*$. The partial derivative for each parameter $\theta[c]$ is:

$$\frac{\partial L_2}{\partial\theta[c]} = \sum_{G\in R}\sum_{t\in G, z: c\leftrightarrow(t,z)} p(z \mid G) - n \sum_{t,z: c\leftrightarrow(t,z)} p_{\text{tuple}}(t, z; \theta),$$

where the first term is due to the linearity and the second term is given by the property $\partial \log A(\theta)/\partial\theta[c] = \sum_{t,z: c\leftrightarrow(t,z)} p_{\text{tuple}}(t, z; \theta)$. Then, we have:

$$\begin{aligned}&\frac{\partial}{\partial\theta[c]} L_2(\theta^*) = 0, \\ &\sum_{G\in R}\sum_{t\in G, z: c\leftrightarrow(t,z)} p(z \mid G) - n \sum_{t,z: c\leftrightarrow(t,z)} p_{\text{tuple}}(t, z; \theta^*) = 0.\end{aligned} \tag{22}$$

It means that the expected counts of the relation on $M \in \mathcal{M}$ equal to the marginal counts of $p_{\text{tuple}}$ at $\theta^*$. Particularly, it also holds for each $c \in M_Z$ since $\mathcal{M}$ contains the marginal $M_Z$. For each possible $z$, it corresponds to exactly one $c \in M_Z$ and Eq. (22) degenerates to:

$$\begin{aligned}\sum_{G\in R} |G| p(z \mid G) &= n \sum_{t} p_{\text{tuple}}(t, z; \theta^*), \\ \sum_{G\in R} |G| p(z \mid G) &= n p_{\text{tuple}}(z; \theta^*).\end{aligned} \tag{23}$$

That is, the expected counts of the relation on $M_Z$ equal to the marginal counts of $p_{\text{tuple}}$ on $M_Z$.

Now, we look into the solution of $\arg\max_\theta L_1$. We have:

$$L_1 = \sum_{G\in R}\sum_{z}\sum_{t\in G} p(z \mid G) \log p_{\text{tuple}}(t \mid z; \theta).$$

By Eq. (14), we have:

$$L_1 = \sum_{G\in R}\sum_{z}\sum_{t\in G} p(z\mid G)\left(\sum_{c\in M\in\mathcal{M}\wedge c\leftrightarrow(t,z)}\theta[c] - \log A'(\theta,z)\right)$$
$$= \sum_{G\in R}\sum_{z}\sum_{t\in G} p(z\mid G)\sum_{c\in M\in\mathcal{M}\wedge c\leftrightarrow(t,z)}\theta[c] - \sum_{G\in R}\sum_{z}\sum_{t\in G} p(z\mid G)\log A'(\theta,z). \quad (24)$$

Similar to $\log A(\theta)$, the log normalization factor $\log A'(\theta,z)$ is convex w.r.t $\theta$ and the first term of Eq. (24) is linear. Therefore, $L_1$ is concave. To show that $\theta^*$ is a solution of $\arg\max_\theta L_1$, it suffices to show that $\partial L_1/\partial\theta[c] = 0$ at $\theta^*$ for all $c$. Using $\partial A'(\theta,z)/\partial\theta[c] = \sum_{t':c\leftrightarrow(t',z)} p_{\text{tuple}}(t'\mid z;\theta)$, we have:

$$\frac{\partial L_1}{\partial\theta[c]} = \sum_{G\in R}\sum_{t\in G, z: c\leftrightarrow(t,z)} p(z\mid G) - \sum_{G\in R}\sum_{z}\sum_{t\in G} p(z\mid G)\sum_{t':c\leftrightarrow(t',z)} p_{\text{tuple}}(t'\mid z;\theta)$$
$$= \sum_{G\in R}\sum_{t\in G, z: c\leftrightarrow(t,z)} p(z\mid G) - \sum_{z}\left(\sum_{G\in R}\sum_{t\in G} p(z\mid G)\right)\left(\sum_{t':c\leftrightarrow(t',z)} p_{\text{tuple}}(t'\mid z;\theta)\right)$$
$$= \sum_{G\in R}\sum_{t\in G, z: c\leftrightarrow(t,z)} p(z\mid G) - \sum_{z}\left(\sum_{G\in R}|G|p(z\mid G)\right)\left(\sum_{t':c\leftrightarrow(t',z)} p_{\text{tuple}}(t'\mid z;\theta)\right). \quad (25)$$

By Eq. (23), we have:

$$\frac{\partial}{\partial\theta[c]}L_1(\theta^*)$$
$$= \sum_{G\in R}\sum_{t\in G, z: c\leftrightarrow(t,z)} p(z\mid G) - \sum_{z} n p_{\text{tuple}}(z;\theta^*)\left(\sum_{t':c\leftrightarrow(t',z)} p_{\text{tuple}}(t'\mid z;\theta^*)\right)$$
$$= \sum_{G\in R}\sum_{t\in G, z: c\leftrightarrow(t,z)} p(z\mid G) - \sum_{z}\sum_{t':c\leftrightarrow(t',z)} n p_{\text{tuple}}(t'\mid z;\theta^*) p_{\text{tuple}}(z;\theta^*)$$
$$= \sum_{G\in R}\sum_{t\in G, z: c\leftrightarrow(t,z)} p(z\mid G) - \sum_{z}\sum_{t':c\leftrightarrow(t',z)} n p_{\text{tuple}}(t',z;\theta^*)$$
$$= \sum_{G\in R}\sum_{t\in G, z: c\leftrightarrow(t,z)} p(z\mid G) - n\sum_{t',z:c\leftrightarrow(t',z)} p_{\text{tuple}}(t',z;\theta^*).$$

By Eq. (22), for each $c$, we have:

$$\frac{\partial}{\partial\theta[c]}L_1(\theta^*) = 0. \quad (26)$$

Therefore, $\theta^*$ is also a solution of $\arg\max_\theta L_1$.

# C BOUNDING PRIVACY CONSUMPTION

In this section, we quantify the privacy consumption of Algorithms 1 and 2, as well as that of Algorithm 6 in [9]. Then, the privacy consumption of Algorithm 3 is the summation of those of its subroutines.

## C.1 Proof of Lemma 6.4

Algorithm 1 has the following queries on private data:

(1) It queries $\text{cnt}^*(z)$ at each EM step in Line 10. $\text{cnt}^*(z)$ is a contingency table of groups and has a sensitivity of $\mu_g$. The total number of queries for $\text{cnt}^*(z)$ is $T$.
(2) It queries $\text{cnt}^*(s,z)$ at each EM step in Line 16. $\text{cnt}^*(s,z)$ is a contingency table of groups and its sensitivity is $\mu_g$. The number of queries for $\text{cnt}^*(s,z)$ is $T$.
(3) It queries latent marginal distributions in Lines 19-22. A latent marginal distribution is a contingency table and the contribution of each group is at most $\tau$. Therefore, its sensitivity is $\tau\mu_g$. As we perform EM for $T$ steps and we have $m_{\mathcal{M}}$ latent marginals, the total number of queries for latent marginals is $m_{\mathcal{M}}T$.

Then, the privacy consumption of Algorithm 1 is:

$$C_1(T, m_{\mathcal{M}}) = T\cdot\mu_g^2\cdot\left(\frac{m_{\mathcal{M}}\cdot\tau^2}{\sigma_l^2} + \frac{1}{\sigma_{size}^2} + \frac{1}{\sigma_z^2}\right).$$

## C.2 Proof of Corollary 6.5

In the case that the tuple multiplier is 1, lemma 2 in [9] shows that the privacy consumption of its Algorithm 6 is:

$$g = \frac{1}{\sigma_{\mathcal{U}}^2} + \frac{2d(d-1)}{\sigma_R^2} + \frac{t \cdot k}{\sigma_h^2} + \frac{(d+t)}{\sigma_o^2}. \tag{27}$$

Notice that its privacy consumption is proportional to their squared sensitivities, i.e. proportional to the squared tuple multiplier. Since our tuple multiplier is at most $\mu_t$, the privacy consumption of the instances of Algorithm 6 [9] in our algorithms is $C_{\text{single}}(d) = \mu_t^2 g$.

## C.3 Proof of Lemma 6.6

We list the privacy consumption of Algorithm 2 as follows:

(1) In Line 1, it invokes the Algorithm 6 of [9] and its privacy consumption is $C_{\text{single}}(d)$.
(2) In Line 7, it uses Algorithm 1 to estimate the parameters when $\mathcal{M}$ contains the initial latent marginals only. Since we have two latent variables as stated in Section 4.6, the number of initial latent marginals is $2d$ and the privacy consumption is $C_1(T, 2d)$
(3) In Line 16, it uses Algorithm 1 to estimate the parameters after the selection of new latent marginals. The privacy consumption is $\sum_{i=1}^{T_C} C_1(1, 2d + i \cdot n_{\text{inc}})$.
(4) In Line 13, it queries $\widetilde{\text{err}}(M')$ for each $M' \in C'$. The total number of the queries is $n_C \cdot T_C$. Since $M'$ is a private contingency table of tuples and $\widetilde{M'}$ is publicly available, the sensitivity of $\widetilde{\text{err}}(M') = \|M' - \widetilde{M'}\|_1$ is $\mu_t$. Then, the privacy consumption is $\mu_t^2 \cdot \frac{n_C \cdot T_C}{\sigma_{\text{err}}^2}$.

Therefore, total privacy consumption is:

$$C_2(\text{FK}(R, R')) = C_{\text{single}}(d) + C_1(T, 2d) + \sum_{i=1}^{T_C} C_1(1, 2d + i \cdot n_{\text{inc}}) + \mu_t^2 \cdot \frac{n_C \cdot T_C}{\sigma_{\text{err}}^2}. \tag{28}$$